%% file: MainFile.tex
\documentclass[lettersize,journal]{IEEEtran}
\usepackage{amsmath,amsfonts}
\usepackage{algorithmic}
\usepackage{algorithm}
\usepackage{array}
\usepackage{textcomp}
\usepackage{stfloats}
\usepackage{url}
\usepackage{verbatim}
\usepackage{graphicx}
\usepackage{cite}

\usepackage{subfig}
\usepackage{longtable}

\usepackage{graphicx}      
\usepackage{amsmath}
\usepackage{amssymb}
\usepackage{arydshln}
\usepackage{afterpage}
\usepackage{xcolor}
\usepackage[export]{adjustbox}

\usepackage{graphicx}

\usepackage{multirow}

\begin{document}

\title{Online set-point estimation for feedback-based traffic control applications 
}

\author{Farzam Tajdari$^{1}$ and Claudio Roncoli$^{1}$
\thanks{$^{1}$Department of Built Environment, School of Engineering, Aalto University, Espoo 02150 Finland
	{\tt\small farzam.tajdari@aalto.fi, claudio.roncoli@aalto.fi}}%
}


\markboth{IEEE TRANSACTIONS ON INTELLIGENT TRANSPORTATION SYSTEMS}%
{Shell \MakeLowercase{\textit{et al.}}: A Sample Article Using IEEEtran.cls for IEEE Journals}


\maketitle

\begin{abstract}                
This paper deals with traffic control at motorway bottlenecks assuming the existence of an unknown, time-varying, Fundamental Diagram (FD). The FD may change over time due to different traffic compositions, e.g., light and heavy vehicles, as well as in the presence of connected and automated vehicles equipped with different technologies at varying penetration rates, leading to inconstant and uncertain driving characteristics.
A novel methodology, based on Model Reference Adaptive Control, is proposed to robustly estimate in real-time the time-varying set-points that maximise the bottleneck throughput, particularly useful when the traffic is regulated via a feedback-based controller. Furthermore, we demonstrate the global asymptotic stability of the proposed controller through a Lyapunov analysis.
The effectiveness of the proposed approach is evaluated via simulation experiments, where the estimator is integrated into a feedback ramp-metering control strategy, employing a second-order multi-lane macroscopic traffic flow model, modified to account for time-varying FDs.
\end{abstract}



\section{Introduction}

\IEEEPARstart{T}{ransport} networks constitute a backbone of our society, enabling mobility of people and distribution of goods. However, due to urbanisation and suboptimal mobility policies and choices, transport infrastructures in and around metropolitan areas are reaching their saturation, with negative effects such as ever-increasing traffic congestion. This causes an increased need for energy, risk of accidents, traffic jams, and driver frustration~\cite{rios2016survey, chhabra2017survey,yang2021information}.
In traffic networks, congestion is typically triggered by the activation of a bottleneck, which occurs when the traffic demand exceeds the road supply. In particular, in a motorway context, whenever there are lane drops, uphills, or curvatures, a bottleneck may appear, which, if activated, may produce a capacity drop, i.e., a reduction of the total discharging flow rate from the bottleneck area, causing travel time delay for the upstream traffic. Traffic congestion then propagates upstream of the bottleneck, until a significant reduction of the demand flow occurs~\cite{chung2007relation,kim2012capacity, hadiuzzaman2013variable, jin2015control}. 

A successful countermeasure able to mitigate or avoid the effects of congestion is traffic control, which consists in using some technological device (e.g., traffic signal, variable message sign, etc.) to regulate the flow entering a specific road area by employing some traffic measurement \cite{ref:Papageorgiou2003}. 
Among other approaches, over the last decades, several feedback-based traffic control methods have been proposed and sometimes implemented, able to partially deal with the aforementioned challenges \cite{Papageorgiou1991,Diakaki2002, carlson2011local, huang2016control, pasquale2020hierarchical, van2019linear, tao2022short, yuan2022selection}.

Despite their design peculiarities, all those control approaches require the knowledge of some features characterising the traffic behaviour in order to work effectively, which include the traffic capacity (i.e., maximum flow able to pass a bottleneck location) and the critical density or occupancy (i.e., the density or occupancy at which capacity occurs).
These quantities are not trivial to obtain or estimate and they require the collection and analysis of traffic data for each area where traffic control is to be applied.
Moreover, even once these parameters are calibrated, they may require constant tuning due to short- and long-term changes in traffic behaviour and characteristics. 
This will be amplified with the appearance of vehicle automation~\cite{ref:Bishop2005}; in fact, it is expected that vehicles with various driving assistance systems, such as Connected and Automated Vehicles (CAVs) are going to co-exist for the next decades, altering the current traffic characteristics and affecting the need for traffic control~\cite{ref:Diakaki2015,ref:Roncoli2015c,ref:Zhang2017,Mattas2018,Papamichail2019}.

A way to deal with this issue is to design and employ adaptive estimation algorithms to automatically tune the parameters (e.g, the set-points) within control strategies. This has been proposed, e.g., in the context of urban traffic control, in~\cite{ref:Kouvelas2011,ref:Kutadinata2016}, where the set-points are tuned on a day-to-day basis.
Papers~\cite{ref:Roncoli2016,tajdari2020feedback} employed a methodology based on discrete-time Extremum Seeking (ES), which is a model-free method, applied to traffic data for real-time optimisation, which has been broadly investigated and utilized in several applications, including, e.g., \cite{ref:Ariyur2003, ref:Kutadinata2016}.
However, even if a set-point is estimated using offline data, it may not always be optimal because of possible changes in traffic behaviour characteristics, which may be caused, by a different traffic composition (e.g., of trucks and cars) or by the presence of CAVs at various penetration rates. 
To our best knowledge, the only existing work dealing simultaneously with control and online set-point estimation is~\cite{yu2021extremum}, which employs a method proposed in \cite{oliveira2016extremum}, developing an online ES control approach to calculate the optimal density input for motorway traffic, when there is a downstream bottleneck. However, such an approach is restricted to a single lane with a one-link network, while the slow convergence speed of the algorithm makes it unsuitable for practical applications.


Apart from the abovementioned approaches, there exist methods capable of simultaneously controlling and identifying the unknown parameters of a system online (see, e.g.,~\cite{slotine1991applied}). 
One suitable method is Model Reference Adaptive Control (MRAC), which is designed to exploit conventional controllers while the controllers' parameters are updated based on model parameters identification, where the model structure is assumed known and parameters values are unknown. 
Such methods have been widely used, e.g., in controlling robotic systems~\cite{jin2018dynamic}, online identification~\cite{gaudio2021parameter}, and noise filtering~\cite{wang2020parameter}. Although employing such adaptive control methods has considerable potential in the domain of traffic control, there is a lack of literature on designing and testing such control schemes.

This paper proposes an adaptive control scheme consisting of a novel globally robust MRAC-based approach for estimating constant or time-varying unknown set-points (in the form of critical densities) for controlling a local motorway bottleneck, with the purpose of maximising the outflow and, consequently, reducing travel delays. 
Our main scientific contributions are as follows.
\begin{itemize}
\item We propose an adaptive dynamic set-point (critical density) estimator, assuming the availability of local traffic measurements, such as the traffic density and flow at the bottleneck.
\item We prove that the estimator is globally asymptotically stable via Lyapunov analysis.
\item We perform numerical investigations employing a state-of-the-art traffic control strategy and non-linear traffic model, to demonstrate the effectiveness of the proposed method. Furthermore, we perform numerical analyses to demonstrate the robustness against parameter choices and disturbances.
\end{itemize}
Note that a preliminary version of this work is included in~\cite{tajdari2021adaptive}, which is extended here in various aspects. First, we provide a more rigorous formulation of the estimation and control problem, while also thoroughly investigating the stability properties of the proposed adaptive estimation law. Second, we redesigned the numerical experiments by considering state-of-the-art modelling and control strategies, while investigating also the robustness of parameter choices.

The paper is structured as follows: the proposed adaptive estimator is described in Section~\ref{sec:estimator}; Section~\ref{sec:experimentsetup} introduces the experiment setup; while in Section~\ref{sec:results} the obtained simulation results are presented; Section~\ref{sec:conclution} concludes the paper, highlighting our main results and indicating future research directions.

\section{Adaptive Estimator Design}
\label{sec:estimator}

\subsection{Preliminary}
We aim at designing an estimator that allows a feedback controller to maximise the outflow at a motorway bottleneck.
As, due to less pronounced fluctuations, it is preferable to employ density as a set-point for the controller~\cite{Papageorgiou2002}, the problem reduces to estimating the critical density at the motorway bottleneck. For this purpose, we first introduce some necessary assumptions; then, we proceed with the controller design; and, finally, we demonstrate the convergence of the estimated values and the stability of the proposed method.

\begin{figure}[tb]
	\centering
	\includegraphics[width= 0.7\linewidth]{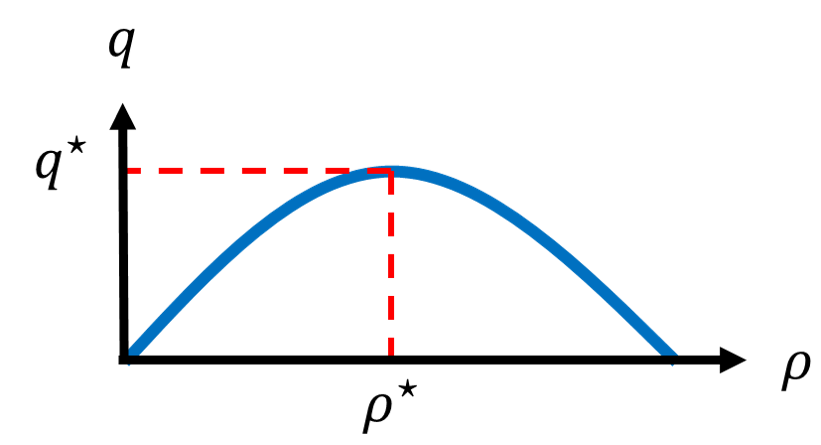}
	\caption{The FD assumed at the bottleneck area.}
	\label{fig:q-r}
\end{figure}

For the design of our estimator, we assume a parabolic flow-density ($q-\rho$) relationship, denoted as fundamental diagram (FD); note however that we will show in Section~\ref{sec:results} that the method is effective also when other shapes for an FD are used, as long as it is concave and has a unique maximum point.
In particular, we employ the following function describing the FD, also depicted in Fig.~\ref{fig:q-r},
\begin{align}
    q = a\rho^2+b\rho,
    \label{eq:parabolic}
\end{align}
where $a$ and $b$ are unknown parameters; function~\eqref{eq:parabolic} has a maximum point ($\rho^{\star}, q^{\star}$) as
\begin{equation}
    \rho^{\star}=\frac{-b}{2a}, \quad q^{\star}=\frac{-b^2}{4a};
\end{equation}
note that $q^{\star}$ is the maximum flow (capacity) and $\rho^{\star}$ is the critical density.

\subsection{Adaptive Estimator design}
By replacing the nominal values of $q^{\star}$ and $\rho^{\star}$ in \eqref{eq:parabolic}, the error of $q$ from $q^{\star}$ is 
\begin{equation}
    q - q^{\star} = a(\rho^2-{\rho^{\star}}^2) + b(\rho-\rho^{\star}).
    \label{eq:errorSys}
\end{equation}
Let us define the integral error states
\begin{align}
    E_q &= \int \left( q - q^{\star} \right) dt \\
    E_{\rho} &= \int \left( \rho - \rho^{\star} \right) dt,
\end{align}
leading to the integral error system
\begin{align}
    \dot{E}_q &= q - q^{\star} \\
    \dot{E}_{\rho} &= \rho - \rho^{\star},
\end{align}
which can be rewritten as
\begin{equation}
        \dot{X} = B_e u_e + r_e,
        \label{eq:integralstate}
\end{equation}
where
\begin{align}
    X &= \left[ \begin{matrix} \int q\;dt \\ \int \rho\;dt \end{matrix} \right], 
    u_e = \left[ \begin{matrix} u_1 \\ u_2 \end{matrix} \right] = \left[ \begin{matrix}  {\rho^{\star}}^2 - \rho^2 \\ \rho - \rho^{\star} \end{matrix} \right]
    \label{eq:estimatorInput}\\
    B_e &= \left[ \begin{matrix} -a&b\\0&1 \end{matrix} \right], 
    r_e = \left[ \begin{matrix} q^{\star}\\\rho^{\star} \end{matrix} \right].
    \label{eq:estimatorMatrix}
\end{align}
We propose to control system \eqref{eq:integralstate} via MRAC~\cite{slotine1991applied}, which allows us to identify the unknown parameters $a$ and $b$ (both appearing in $B_e$) while minimising the tracking error simultaneously.
In order to proceed, we introduce the feedback control law
\begin{equation}
    u_e = -\hat{B}X-\hat{C}r_e,
    \label{eq:controlLaw}
\end{equation}
where $\hat{B}$ and $\hat{C}$ are unknown matrices that need to be estimated. Defining $\hat{\Pi} = \left[ \begin{matrix} -\hat{B} & -\hat{C} \end{matrix} \right]$, \eqref{eq:controlLaw} becomes
\begin{equation}
    u_e = \hat{\Pi} \left[ \begin{matrix} X \\ r_e \end{matrix} \right].
    \label{eq:u_eOld}
\end{equation}
We then introduce a model reference
\begin{equation}
    \dot{X}_M = -A_M X_M + B_M r_e,
    \label{eq:referenceModel}
\end{equation}
where $A_M$ and $B_M$ are arbitrarily defined matrices that make the dynamic of model reference stable.
Let us define the error between the integral states and the model reference $e=X-X_M$, whose dynamics is defined as
\begin{align}
    &\dot{e} = \dot{X} - \dot{X}_M \nonumber \\
    &\!\!= \!B_e (-\hat{B}X\!-\!\hat{C}r_e)\!+\!r_e\!+\!A_M X_M\!-\!B_M r_e\!+\!A_M X\!-\!A_M X  \nonumber \\
    &\!\!= \!-A_M (X-X_M)\!+\!B_e(-\hat{B}\!+\!A_M)X\!+\!B_e(-\hat{C}\!-\!\frac{B_M\!-\!I}{B_e}) r_e\nonumber \\
    &\!\!= \!-A_M e + B_e(-\hat{B}+A_M)X \!+\! B_e(-\hat{C} \!-\! \frac{B_M \!-\! I}{B_e}) r_e,
    \label{eq:errorXXM}
\end{align}
which, converting to the Laplace domain, leads to
\begin{align}
    &\!\!e =\frac{B_e}{sI \!+\! A_M} \left[ \begin{array}{c:c} -\hat{B}+\frac{A_M}{B_e}\; & \;-\hat{C}-\frac{B_M-I}{B_e} \end{array} \right] \left[ \begin{matrix} X \\ r_e \end{matrix} \right] \nonumber \\
    &\!\!\!=\frac{|B_e|}{sI \!+\! A_M}\!\!\left[\!\!\! \begin{array}{c:c} \!-\hat{B}\Theta(B_e)\!+\!\frac{A_M}{|B_e|}\!\! & \!\!-\hat{C}\Theta(B_e)\!-\!\frac{B_M-I}{|B_e|} \end{array}\!\!\!\right] \!\!\left[ \begin{matrix} X \\ r_e \end{matrix} \right]\!\!,
    \label{eq:errorDynamic}
\end{align}
where $s$ is the Laplace variable, and $\Theta$ is a sign operator defined as follows
\begin{align}
\Theta(i) &= 
\begin{cases}
1,  & \textrm{if } i>0 \\
0,  & \textrm{if } i=0\\
-1, & \textrm{if } i<0.
\end{cases} 
\end{align}
Accordingly, the error dynamic of~\eqref{eq:errorDynamic} is stable over time, as $A_M$ is chosen as a stable matrix, if changes of $\hat{\Pi}$ are restricted to a finite domain or $\hat{\Pi}$ is converging to a certain value. 
As in \eqref{eq:errorDynamic}, $\Theta(B_e)$ appeared, we rewrite~\eqref{eq:u_eOld} to facilitate the calculation as 
\begin{equation}
    u_e = \Theta(B_e)\hat{\Pi}\left[ \begin{matrix} X \\ r_e \end{matrix} \right].
    \label{eq:Usign}
\end{equation}
Then, by replacing~\eqref{eq:Usign} into~\eqref{eq:integralstate}, we obtain
\begin{align}
    \dot{X} = B_e \Theta(B_e)\left[ \begin{matrix} -\hat{B} & -\hat{C} \end{matrix} \right] \left[ \begin{matrix} X \\ r_e \end{matrix} \right] + r_e,
\end{align}
which, defining $|B_e| = B_e \Theta(B_e)$, results in
\begin{align}
    \dot{X} = -|B_e|\hat{B}X - \left( |B_e| \hat{C}-I\right)r_e;
\end{align}
this system is exponentially stable around $r_e$ if $\lim_{t\to\infty} \hat{B}~\to~|B_e|$ and $\hat{C}~\to~\frac{1}{|B_e|}$, as the matrix $|B_e|\hat{B}$ is a Hurwitz matrix proves the exponential stability (see Chapter 3 of~\cite{khalil2015nonlinear}).

In order to investigate the convergence of $\hat{\Pi}$, the following Lyapunov function is used:
\begin{equation}
    \mathcal{V} = X^T \mathcal{P} X
+ \hat{\Pi}^T \Gamma^{-1} \hat{\Pi},
\label{eq:Lyapunov}
\end{equation}
where $\mathcal{P}\geq 0$ and $\Gamma > 0$ imply that $\mathcal{V} > 0$. 
In order to guarantee stability, it would be enough if $\dot{\mathcal{V}} \leq 0$, then
\begin{equation}
    \frac{d\mathcal{V}}{dt} = \dot{X}^T \mathcal{P}X + X^T\mathcal{P}\dot{X} + \dot{\hat{\Pi}}^T\Gamma^{-1}\hat{\Pi} + \hat{\Pi}^T \Gamma^{-1} \dot{\hat{\Pi}}.
\end{equation}
By replacing $v = \left[ \begin{matrix} X \\ r_e \end{matrix} \right]$ in~\eqref{eq:Usign}, we obtain $u_e = \Theta(B_e) v^T \hat{\Pi}$, thus
\begin{equation}
    \frac{d\mathcal{V}}{dt} = 2X^T\mathcal{P}B_e \Theta(B_e) v^T \hat{\Pi} + 2\dot{\hat{\Pi}}^T\Gamma^{-1}\hat{\Pi}.
    \label{eq:dvdt}
\end{equation}
If we define $\mathcal{P} B_e = C^T$, $e = X^{T} C^{T}$, and $\frac{d \mathcal{V}}{dt} = 0$, then from~\eqref{eq:dvdt} we have
\begin{equation}
    -2 \; \Theta(B_e)ev^T \hat{\Pi} = 2\dot{\hat{\Pi}} \; \Gamma^{-1} \; \hat{\Pi};
\end{equation}
thus, a sufficient condition for stability is that the changes in the unknown parameters are
\begin{equation}
    \dot{\hat{\Pi}} = -\Gamma \; \Theta(B_e) v e^T,
    \label{eq:adaptiveLaw}
\end{equation}
where $\Gamma$ is known as the \emph{growth rate} of the estimation law. For our problem, and according to~\eqref{eq:estimatorMatrix}, as the unknown components of $B_e$ ($a$ and $b$) are positive (while the known components 0 and 1 are fixed), $\Theta(B_e)$ has no impact on the performance of the estimator and can be neglected, resulting in
\begin{equation}
    \dot{\hat{\Pi}} = -\Gamma \; v e^T.
    \label{eq:adaptiveLawSimple}
\end{equation}

\subsection{The growth rate of the estimator}
Although any positive constant value for $\Gamma$ is theoretically sufficient to guarantee the globally asymptotically stability of the estimator, the degree of robustness may define the converging time, which is an important property in practical implementations. 
In particular, we observed in numerical experiments that, when $|\dot{\hat{\Pi}}|>1$, the estimated values of $q^{\star}$, and $\rho^{\star}$ feature high oscillations, resulting in a deterioration of the controller performance.
Thus we redefine $\Gamma$ as a time-varying parameter as follows:
\begin{equation}
    \Gamma(t) = \left( \int_{0}^{t} v(r) v^T(r) dr \right)^{-1}, 
    \label{eq:GammaOrigin}
\end{equation}
where
\begin{equation}
    \frac{d}{dt}\left( \Gamma^{-1}(t)\right) = v(t) v^T(t).
    \label{eq:integralGamma}
\end{equation}
From \eqref{eq:adaptiveLawSimple} we have that
\begin{align}
    \dot{\hat{\Pi}} &= -\Gamma(t) \; v(t) e^T(t) \label{eq:EstimationLawGamma}\\
    \Gamma^{-1}(t) \; \dot{\hat{\Pi}} + v(t) v^T(t)\; \hat{\Pi} &= v(t) v^T(t) \Pi \\
    \frac{d}{dt}(\Gamma^{-1}(t) \hat{\Pi}) &= v(t) v^T(t) \Pi\\
    \frac{d}{dt}(\int_{0}^{t} v(r) v^T(r) dr \hat{\Pi}) &= v(t) v^T(t) \Pi\\
    \int (v^T(r) \Pi - v^T(r) \hat{\Pi}) dr &= 0.
    \label{eq:lease_squareResult}
\end{align}
Essentially, \eqref{eq:lease_squareResult} implies that the proposed estimation law is minimizing a cost function based on the well-known least-square method, as follows
\begin{equation}
    J = \int_{0}^{t} ||v^T(r) \hat{\Pi} - v^T(r) a||^2 dr.
    \label{eq:least-square}
\end{equation}
However, while implementing  the estimator, it is desirable to update the gain $\Gamma(t)$
directly, rather than using \eqref{eq:integralGamma} and then inverting the matrix $\Gamma^{-1}$, which may cause numerical issues. Instead, by using the identity matrix
\begin{equation}
    \frac{d}{dt}\left( \Gamma(t) \Gamma^{-1}(t) \right) = \dot{\Gamma}(t) \Gamma^{-1}(t) + \Gamma(t) \frac{d}{dt}\left( \Gamma^{-1}(t) \right) = 0,
\end{equation}
we obtain
\begin{equation}
    \dot{\Gamma}(t) = - \Gamma(t) v(t)v^{T}(t) \Gamma(t).
    \label{eq:GammaUpdate}
\end{equation}

While using \eqref{eq:EstimationLawGamma} and \eqref{eq:GammaUpdate} for online estimation, we need to specify initial values for the estimated parameters and the gain of growth rate. However, the initialisation may be challenging, as from \eqref{eq:EstimationLawGamma} and \eqref{eq:GammaUpdate} it results that $\Gamma(0)$ should be a very large value (theoretically approaching infinity), whereas $\hat{\Pi}$ is initially undefined. To tackle this challenge, assuming proper finite values to initialize $\Gamma$ and $\hat{\Pi}$ can be a problem solver. One should use the best guess to initialize $\hat{\Pi}$, a proper initial value of the gain $\Gamma(0)$ must be opted as high as allowed by the noise sensitivity extracted from the dynamic of the system analysis. Note that, for the sake of simplicity, $\Gamma(0)$ may be chosen as a diagonal matrix.



\subsection{Parameter convergence} 
Theoretically, the convergence properties of the estimator can be revealed via solving the differential equations~\eqref{eq:integralGamma} and \eqref{eq:EstimationLawGamma}, assuming the absence of noise and parameters variation. From \eqref{eq:integralGamma}, \eqref{eq:EstimationLawGamma}, and \eqref{eq:GammaUpdate}, one may see that
\begin{align}
    &\Gamma^{-1}(t) = \Gamma^{-1}(0) + \int_{0}^{t} v(r)v^{T}(r) dr \label{eq:condition1}\\
    &\frac{d}{dt}\left( \Gamma^{-1}(t) \Tilde{\Pi}(t)\right) = 0,
    \label{eq:condition2}
\end{align}
where $\Tilde{\Pi} = \hat{\Pi} - \Pi$; thus, assuming equilibrium conditions (i.e., $\dot{\Pi} = 0$), we obtain
\begin{equation}
    \tilde{\Pi}(t) = \Gamma(t) \Gamma^{-1}(0) \tilde{\Pi}(0).
    \label{eq:condition3}
\end{equation}
If $v$ is such that 
\begin{equation}
    \lim_{t \to \infty} \lambda_{\textrm{min}}\left( \int_{0}^{t} v(r) v^{T}(r) dr \right) \rightarrow \infty, 
    \label{eq:convergance}
\end{equation}
where $\lambda_{\textrm{min}}(\cdot)$ denotes the smallest eigenvalue of its argument, then the gain matrix
converges to zero and the estimated parameters asymptotically (although usually not
exponentially) converge to the true parameters. Indeed,
for any positive integer $k$,
\begin{equation}
    \int_{0}^{k\sigma + \sigma} \!\!\! v(r)v^{T}(r) dr = \sum_{i = 0}^{k} \int_{i\sigma}^{i\sigma + \sigma} \!\!\! v(r)v^{T}(r) dr \geq k \alpha_1 I.
    \label{eq:stabilityGammaRule}
\end{equation}
Thus, if $v$ is persistently excited, \eqref{eq:stabilityGammaRule} is satisfied; then, according to \cite{anderson1977exponential, morgan1977uniform}, $\Gamma \rightarrow 0$ and $\tilde{\Pi} \rightarrow 0$.

Note that the impact of the initial gain value and the initial parameter value on the estimation process is observable from \eqref{eq:condition1}, \eqref{eq:condition2}, and \eqref{eq:condition3}. In fact, a small error in the parameter's initialisation value ($\tilde{\Pi}(0)$), always leads to a small parameter estimation error. Whereas, a large initial gain $\Gamma(0)$ results in a small parameter estimation error. Generally, based on \eqref{eq:GammaOrigin}, $\Gamma$ is naturally a very small value and according to \eqref{eq:GammaUpdate}, it is exponentially converging to zero. Thus, if $\Gamma(0)$ is not big enough we may have no update or very low-speed update in the parameter estimation. 
This is more evident if we select $\Gamma(0) = \Gamma_{0}I$, which results in
\begin{equation}
    \tilde{\Pi} = \left( I + \Gamma_{0} \int_{0}^{t} v(r)v^{T}(r) dr \right)^{-1} \tilde{\Pi}(0).
\end{equation}

\subsection{Robustness to noise in the density and flow measurement}
Generally, the least-squares method \eqref{eq:least-square} used for the designed growth rate ($\Gamma$) calculated via \eqref{eq:GammaUpdate} performs robustly with respect to noise and disturbance. Proper noise-rejection capability results from the fact that noise, particularly if characterised by high frequency, is averaged out. The estimator's inability in tracking highly fluctuating parameters (different from switching parameters' values) is also understandable intuitively, from two different viewpoints. In mathematical terms, $\Gamma(t)$ converges to zero when $v$ is persistently excited according to \eqref{eq:condition2}, i.e., the parameter update is essentially shut off after some time, and the changing parameters cannot be updated anymore. In practical terms, the
least-square estimator tries to fit all the data up to the current time, while, in practice, the previous data is extracted by the previous parameters.

\subsection{The estimator framework}
\begin{figure*}[tb]
	\centering
	\includegraphics[width= 0.85\linewidth]{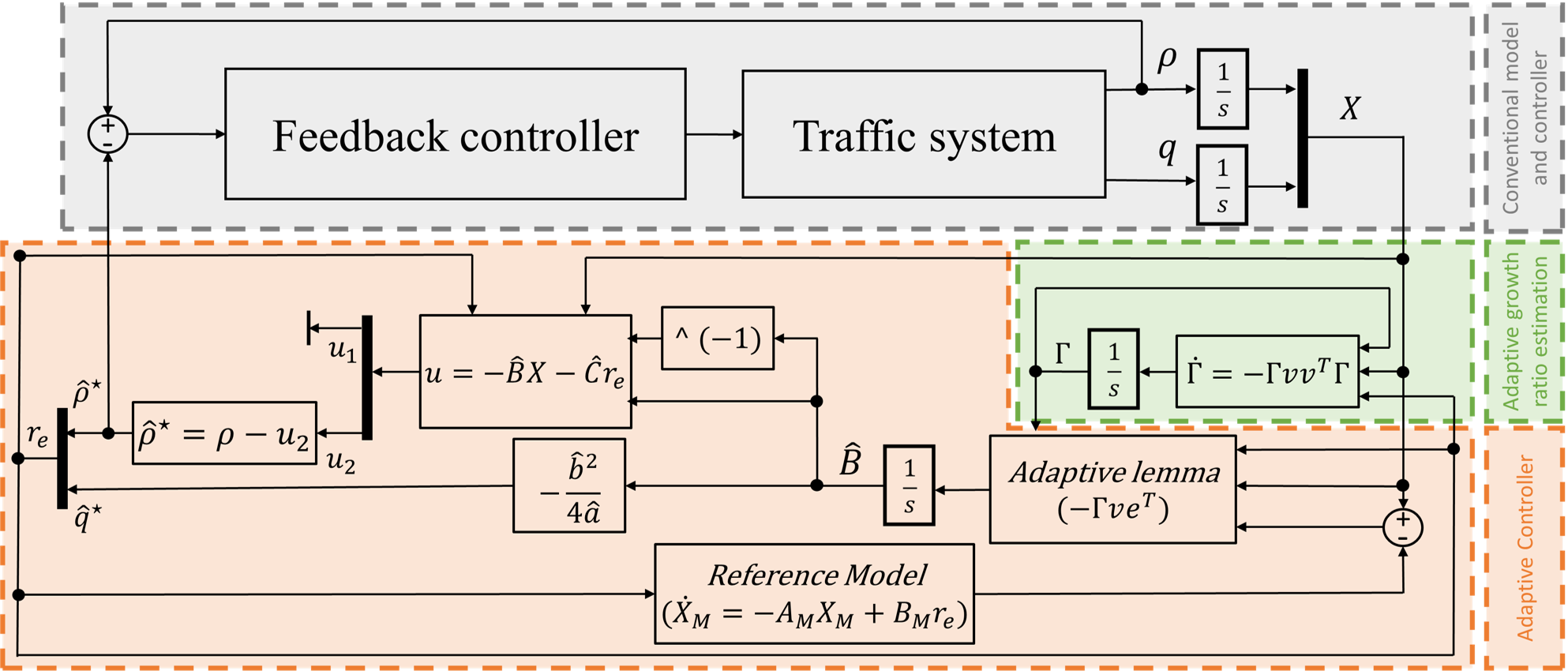}
	\caption{Closed-loop diagram of the controller-estimator integration.}
	\label{fig:diagram}
\end{figure*}

The overall framework proposed in this work, depicted in Fig.~\ref{fig:diagram}, consists of a feedback controller designed to maintain the density at a motorway bottleneck around the critical set-point estimated via the proposed methodology. 
The framework is composed of three main parts: a) the feedback traffic control loop (grey part); b) the adaptive estimator (orange part), and c) the adaptive estimator's parameters (grow ratio) estimation (green part). The grey part essentially includes any feedback controller that utilises density as a set-point to maximise bottleneck throughput. The orange part represents the estimation process of $\rho^{\star}$ and $q^{\star}$, while the growth ratio of the estimator ($\Gamma$) is adaptively estimated as shown in the green part. 

To implement the parameter estimation in discrete form, we consider $\dot{\hat{\Pi}}(k) = \frac{\hat{\Pi}(k) - \hat{\Pi}(k-1)}{\Delta t}$ and $\dot{\Gamma}(k) = \frac{\Gamma(k) - \Gamma(k-1)}{\Delta t}$; then, the adaptation rule~\eqref{eq:adaptiveLawSimple} and the gain update~\eqref{eq:GammaUpdate} turn into
\begin{align}
    \!\!\!\!\!\hat{\Pi}(k) \!&=\! \hat{\Pi}(k\!-\!1) -\Delta t(\Gamma(k\!-\!1) \; v(k\!-\!1) e(k\!-\!1)^T), \label{eq:a_adaptiveLawDis} \\
    \!\!\!\!\!\Gamma(k) \!&=\! \Gamma(k\!-\!1) \!-\! \Delta t(\Gamma(k\!-\!1)v(k\!-\!1)v^T\!(k\!-\!1)\Gamma(k\!-\!1)). 
    \label{eq:adaptiveLawDis}
\end{align}

The estimates~$\hat{\rho}^{\star}$, namely the set-point for the feedback controller, can be obtained as (see~\eqref{eq:estimatorInput})
\begin{equation}
    \hat{\rho}^{\star}(k)=\rho(k)-u_2 (k),
    \label{eq:rhoStar}
\end{equation}
where $u_2$ is an element of $u_e$, which can be computed from~$\hat{\Pi}$ and measured variables~$\rho$. Moreover, the estimates for the maximum outflow~$\hat{q}^{\star}$ is calculated as (see~\eqref{eq:estimatorInput}) 
\begin{equation}
    \hat{q}^{\star}(k) = -\frac{\hat{B}_{1,2}^2}{4 \hat{B}_{1,1}}.
    \label{eq:qStar}
\end{equation}
Note that \eqref{eq:rhoStar} and \eqref{eq:qStar} are defined so that we avoid having dependent parameter estimation, which is necessary to achieve convergence to the true values (see \cite{slotine1991applied}).

\section{Experimental Set-Up} \label{sec:experimentsetup}
We now proceed by demonstrating via numerical experiments the effectiveness of the proposed methodology.
We firstly introduce the traffic simulation model and the feedback control ramp metering strategy considered in our experiments, followed by the evaluated scenarios and the parameters utilised for the model, controller, and estimator.  

\subsection{The macroscopic traffic flow model METANET}

The macroscopic traffic flow model METANET \cite{ref:Messmer1990} is selected for the numerical experiments. 
METANET is a second-order traffic flow model consisting of two interconnected dynamic equations, which describe the evolution of traffic density and (space) mean speed, respectively.
To define a space-time discretized model, the considered freeway stretch is subdivided into $N$ cells of lengths $L_i$, $i = 1, 2, \ldots, N$; whereas the time $t = kT$ is discretized, where $T$ is the simulation time step and $k = 0, 1, \ldots$ is the discrete-time index. 
The traffic characteristics of each cell are macroscopically identified by the following traffic variables:
\begin{itemize}
    \item traffic density~$\rho_i(k)$, as the number of vehicles in cell~$i$ at time~$t = kT$,
divided by $L_i$ and by the number of lanes~$\lambda_i$ in the considered cell (measured in veh/km/lane);
    \item  mean speed~$v_i(k)$ as the mean speed of vehicles in cell~$i$ at time~$t = kT$ (measured in km/h);
    \item  traffic flow~$q_i(k)$ as the number of vehicles leaving cell~$i$ during the time period~$(kT, (k+1)T ]$, divided by~$T$ (measured in veh/h).
\end{itemize}
The equations of the second-order macroscopic traffic flow model used to calculate the traffic variables are:
\begin{align}
\label{eq:model1}
    \rho_i(k\!+\!1) &= \rho_i(k) \!+\!\! \frac{T}{L_i \lambda_i}\! \left[ q_{i-1}(k) \!\!-\! q_i(k) \!\!+\! r_i(k) \!\!-\! s_i(k) \right] \\
    v_i(k\!+\!1) &= v_i(k) + \frac{T}{\tau} \left[ V \left( \rho_i(k) \right) - v_i(k) \right] \nonumber \\
    &\quad + \frac{T}{L_i} v_i(k) \left[ v_{i-1}(k) - v_i(k) \right] \nonumber \\
    &\quad - \frac{\nu T}{\tau L_i} \frac{\rho_{i+1}(k) - \rho_i(k)}{\rho_i(k) + \kappa} \!-\! \frac{\delta T}{L_i \lambda_i} \frac{r_i(k) v_i(k)}{\rho_i(k) + \kappa} \\
    q_i(k) &= \rho_i(k) v_i(k) \lambda_i \\
    V \! \left( \rho_i(k) \right) &= v^{\textrm{free}}_i(k) \; \textrm{exp} \left[ -\frac{1}{\alpha_i(k)} \left( \frac{\rho_i(k)}{\rho^{\textrm{cr}}_i(k)} \right)^{\alpha_i(k)} \right], 
    \label{eq:modelend}
\end{align}
where $\tau$ (time constant), $\nu$ (anticipation constant), $\kappa$ (model parameter) are global parameters given for the whole motorway; $ r_i(k)$ and $s_i(k)$ are the on-ramp inflow and off-ramp outflow, respectively; $V \left( \rho_i(k) \right)$ is a speed–density relationship that represents the FD; finally, $v^{\textrm{free}}_i(k)$ (free-flow speed), $\rho^{\textrm{cr}}_i(k)$ (critical density), and $\alpha_i(k)$ (model parameter) are parameters that characterise the FD in each cell, which, differently from the original formulation, in this work are considered time-dependent to describe the possibility of the FD to change over time.
METANET is widely considered one of the most accurate macroscopic traffic models, capable of reproducing traffic instabilities and the capacity drop effect, which are essential for evaluating traffic control strategies.

\subsection{ALINEA Ramp metering strategy}
We assume traffic is controlled by the well-known ramp-metering feedback controller ALINEA \cite{Papageorgiou1991}. 
The controller ALINEA is designed to maintain the total (cross-lane) density at its critical value in the bottleneck segment, which, in turn, is expected to maximise the bottleneck throughput.
This is done by manipulating the ramp inflow via an I-type controller, according to the following control law
\begin{equation}
\label{eq:alinea}
u(k)=u(k-1)+K_{A} \left( \hat{\rho}^{\star}_{\hat{i}}(k) - \rho_{\hat{i}}(k)\right), 
\end{equation}
where $u(k)$ is the controlled input (ramp flow); $\rho_{\hat{i}}(k)$ is the (measured) density at bottleneck cell $\hat{i}$; $\hat{\rho}^{\star}_{\hat{i}}(k)$ is the estimated set-point for the density at the bottleneck cell; and $K_A$ is the controller gain, which can be defined, e.g., via a trial-and-error procedure.
Note that, due to input saturation, the value $u(k-1)$ used in the right-hand side of \eqref{eq:alinea}
should be the bounded value of the previous time step, i.e., after
application of the upper and lower bounds constraints (considering, e.g., $u^{\textrm{min}}$ and $u^{\textrm{max}}$ as the lower and upper bound, respectively, for the input $u(k)$) in order to avoid
the wind-up phenomenon in the regulator.




Since ramp metering actions may create a queue outside the motorway network, we introduce the following dynamics for the (vertical) queue length $w(k)$ (in veh)
\begin{equation}
\label{eq:w}
w(k+1)=w(k)+T\left(d(k)-u(k)\right),
\end{equation}
where $d(k)$ is the on-ramp external demand during time interval~$ (k,k+1] $. 

In addition, in the presented experiments (as well as in the majority of real-life situations), we assume that the ramp capacity is smaller than the mainstream one; in the opposite case, there may be a need to consider the presence of on-ramp queues also for the no-control case, thus $d$ should also be saturated.


\subsection{Network description and simulation configuration} \label{sec:network}

\begin{figure}[tb]
	\centering
	\includegraphics[width=\linewidth]{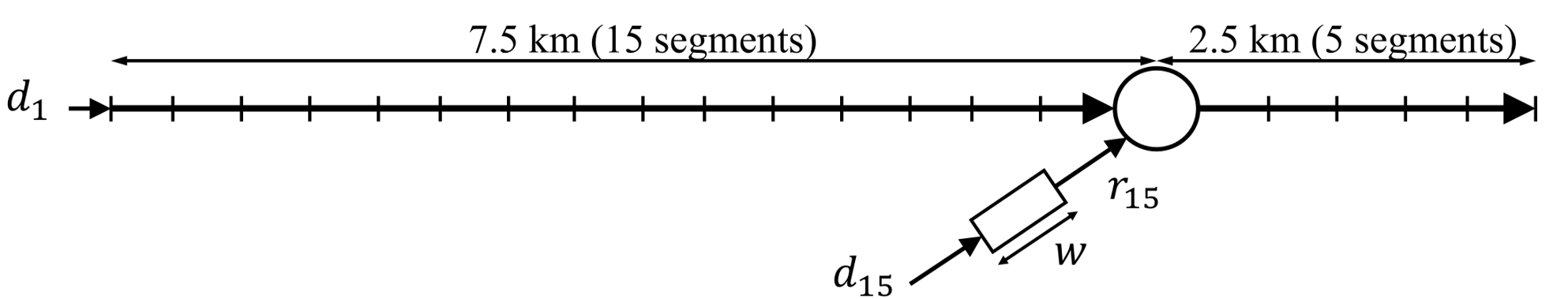}
	\caption{The motorway stretch utilised in the simulation experiments.}
	\label{fig:figRamp10}
\end{figure}

We consider a two-lane motorway stretch, depicted in Fig.~\ref{fig:figRamp10}, which contains a metered on-ramp to test and evaluate the performance of the proposed strategy in presence of changing FD. 
The stretch considered contains two origins, i.e., the main-stream and an
on-ramp, two freeway links, and one destination.
In particular, we consider a network 
composed of 20 segments of the same length $L_i=0.5 \textrm{ km}$, while we employ a time step $T=10 \textrm{ s}$. 
The simulation horizon is 4~h, corresponding to $K=1440$~steps.

We assume that the FD changes from FD$_1$ to FD$_2$ in the middle of our simulations (i.e., after 2~h, $k=720$), which may reflect different traffic compositions (e.g., a high number of heavy vehicles altering the traffic characteristic of the motorway). 
We employ typical METANET parameters from~\cite{ref:Messmer1990}, which are shown in Table \ref{tab:parFD}.

\begin{figure}[tb]
	\centering
	    \subfloat{\includegraphics[width = 0.8\linewidth]{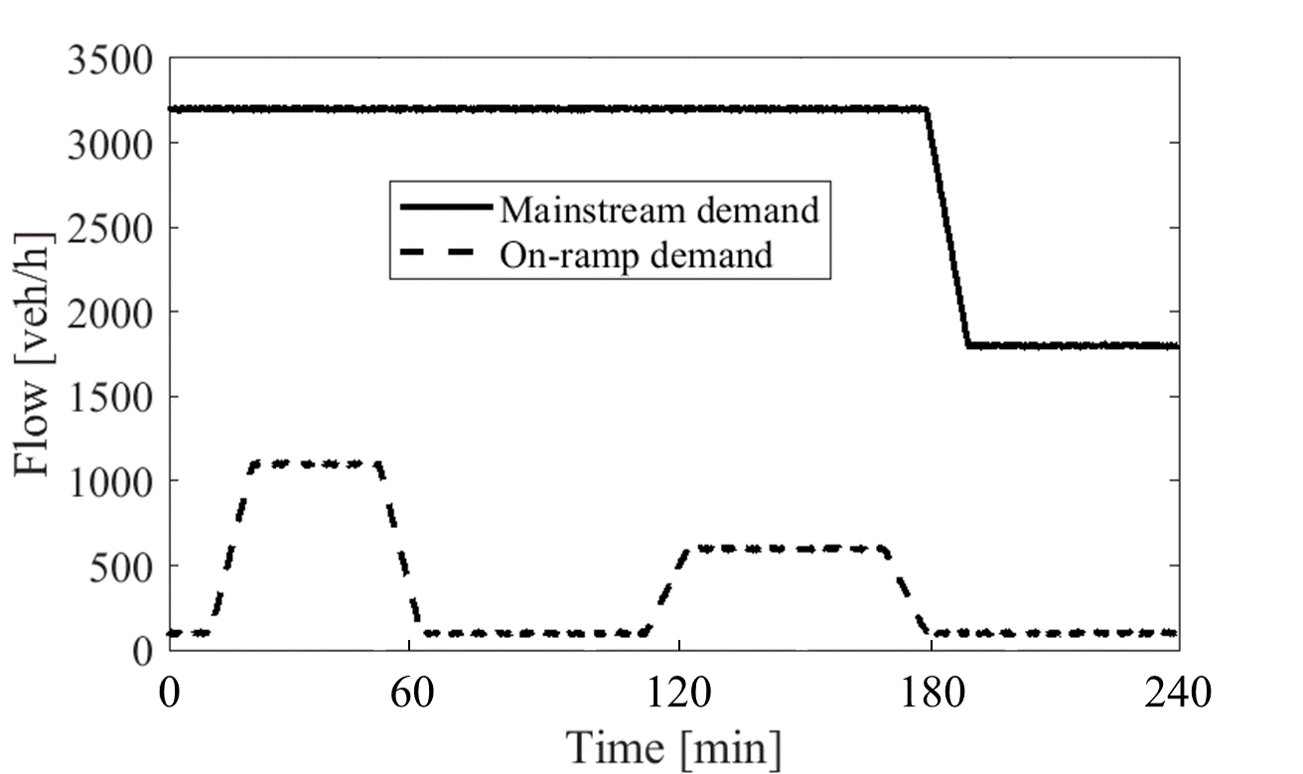}}\hspace{0 cm}
	\caption{Traffic demand used in the simulation experiments.}
	\label{fig:demand}
\end{figure}

To examine the effects of the time-varying FD and the potential of ramp metering to mitigate congestion, we consider the following demand scenario (see also Fig.~\ref{fig:demand}). The mainstream demand is kept constant at a relatively high level (about 80\% of the nominal capacity) for the first 3~h of simulation, dropping to a low level (less than 50\% of the nominal capacity) during the last hour; the latter being a cool-down period useful for ensuring that any congestion dissipates to allow fair numerical comparisons.
The demand on the on-ramp increases for the first time after 10 min to a high value, remains constant for 30 min, and decreases to a constant low value. This is expected to create some congestion while traffic behaves according to FD$_1$. Then, after the FD changes to FD$_2$, the on-ramp demand increases for a second time, remains constant for 45 min, and finally decreases to a constant low value. This scenario is defined such that two independent congestion instances occur with different FDs.

\begin{table}[tb]
	\renewcommand{\arraystretch}{1.3}
	\caption{Parameters used in the nonlinear multi-lane traffic flow model.}
	\label{tab:parFD}
	\centering
	\input{tabParFD}
\end{table}

We employ as a performance metric the Total Time Spent (TTS) over a finite time horizon $K$, defined as
\begin{equation}
\textrm{TTS} = T \sum_{k=0}^{K} \sum_{i=0}^{N} L_i  \rho_{i}(k) + T w(k),
\label{eq:tts}
\end{equation}
which allows to consider both the effects of congestion created in the mainstream and the queue generated at the on-ramp when ramp metering is implemented.

\subsection{Reference model formulation}
\label{subsec:ReferenceModel}

As discussed in Section~\ref{sec:estimator}, the proposed estimator requires the definition of a reference model characterised by stable dynamics, where one of the states is the integral of the other state. Here, we employ the well-known mass-spring-damper model~\cite{armaghan2011design}, which is a two-state system with globally stable dynamics. Actually, for each of the estimated $\hat{q}^{\star}$, or $\hat{\rho}^{\star}$ we are using an independent mass-spring-damper model; thus, describing them in a single system, we employ a four-state system with stable states around $r_e$, defined as follows
\begin{align}
    \dot{X}_r &= A_r X_r + B_r r_e \\
    A_r &\!=\! \left[ \begin{matrix} 0&1 & 0 & 0\\-K_r&-C_r & 0 & 0 \\ 0 & 0 & 0 & 1\\0 & 0 & -K_r&-C_r \end{matrix}  \right], B_r \!=\! \left[ \begin{matrix} 0 & 0\\K_r & 0 \\ 0 & 0\\0 & K_r  \end{matrix}\right],
\end{align}
where $X_r := \left[ \begin{matrix} \int \hat{q}^{\star}\;dt, & \hat{q}^{\star}, & \int \hat{\rho}^{\star}\;dt, & \hat{\rho}^{\star} \end{matrix}  \right]^T$, thus $X_M = \left[ \begin{matrix} X_{r_{1,1}} & X_{r_{3,1}}\end{matrix}  \right]^T$ as $X_M \equiv X$ in \eqref{eq:errorXXM}, and $K_r > 0$ and $C_r > 0$ are the spring and damper coefficients, respectively. The system is globally stable to $r_e$ as all the eigenvalues of $A_{r}$ are negative and the pair ($A_{r}, B_{r}$) is stabilisable (see, e.g.,~\cite{ref:Williams2007}).

In the performed experiments, while applying control, we employ the feedback law \eqref{eq:alinea} in the model (\ref{eq:model1})-(\ref{eq:modelend}). The controller gain is set as~$K_A~=~15$ (tuned via trial-and-error), while the set-point is determined via \eqref{eq:rhoStar} iteratively calculating~\eqref{eq:a_adaptiveLawDis}.  
A sensitivity analysis involving parameters $K_r$ and $C_r$ has been carried out, which is reported in Section~\ref{subsec:Sensitivity} (see also~\cite{tajdari2020feedback}); for most of the experiments we use the following values: $K_r=10$ and $C_r=2$.
Finally, the initial value of the growth rate in~\eqref{eq:adaptiveLawDis} is set as $\Gamma(0) = 20 $.

\section{Experimental Results} 
\label{sec:results}
We now proceed by presenting quantitative results demonstrating the performance of the proposed methodology for different settings of the proposed estimator. 
We define and consider the following baseline scenarios for our comparisons. 
\begin{itemize}
    \item \textbf{Scenario 1:} the no-control case, where the ramp flow is not metered, therefore congestion is expected to be formed;
\end{itemize}
\begin{itemize}
    \item \textbf{Scenario 2:} a controlled case with known set-points, where ramp metering is active, considering that critical densities (thus, the set-points) are perfectly known (obtained, e.g., by analysing the no-control case results);
\end{itemize}
\begin{itemize}
    \item \textbf{Scenario 3:} a controlled case where the set-point is maintained constant during the whole simulation, where the set-point is set equal to the critical density defined for the first half of simulation (Scenario~3-a) and equal to the critical density defined for the second half of simulation (Scenario~3-b).
\end{itemize}
We reasonably expect that the no-control case (Scenario~1) is a lower bound for performance, while the controlled case with known set-points (Scenario~2) is an upper bound for the improvements that may be achieved.
We then implement and evaluate controlled scenarios utilising our estimator as follows. 
\begin{itemize}
    \item \textbf{Scenario~4:} we test the estimator by setting as initial set-point the critical densities of FD$_1$ and FD$_2$, in Scenarios~4-a and 4-b, respectively;
\end{itemize}
\begin{itemize}
    \item \textbf{Scenario~5:} we test our estimator by considering initial set-points values that are very high (Scenario~5-a) and very low (Scenario~5-b) compared to the actual ones. 
\end{itemize}
In the plots presented afterwards, we use the blue colour for reporting the results of the FD employed during the first half of the simulation (FD$_1$) and the red colour for the FD employed during the second half of the simulation (FD$_2$).

\subsection{Scenario 1: No-control case} 
\label{sec:no-control}
\begin{figure}[tb]
		\centering
    \subfloat[]{\includegraphics[width = 0.49\linewidth]{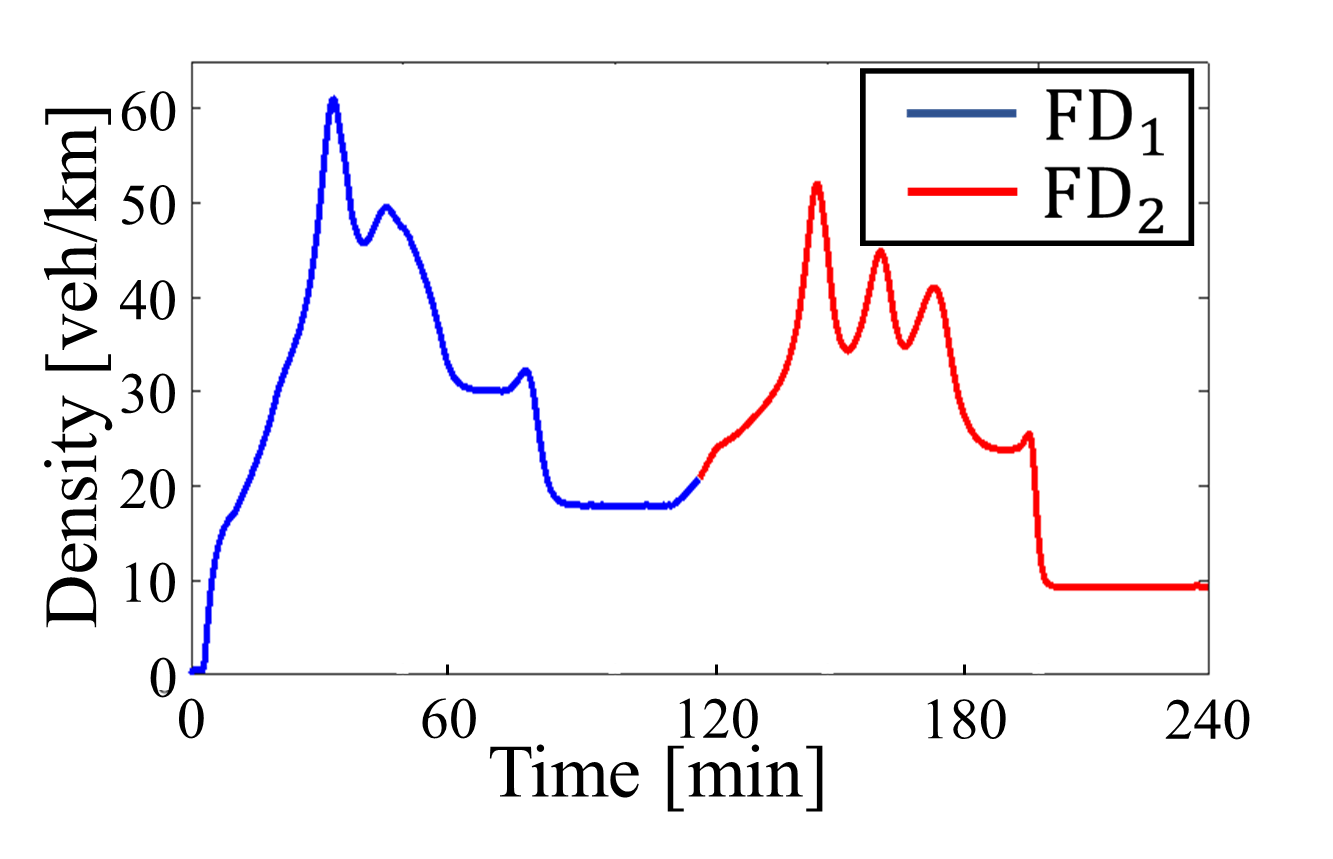}}
    \subfloat[]{\includegraphics[width = 0.49\linewidth]{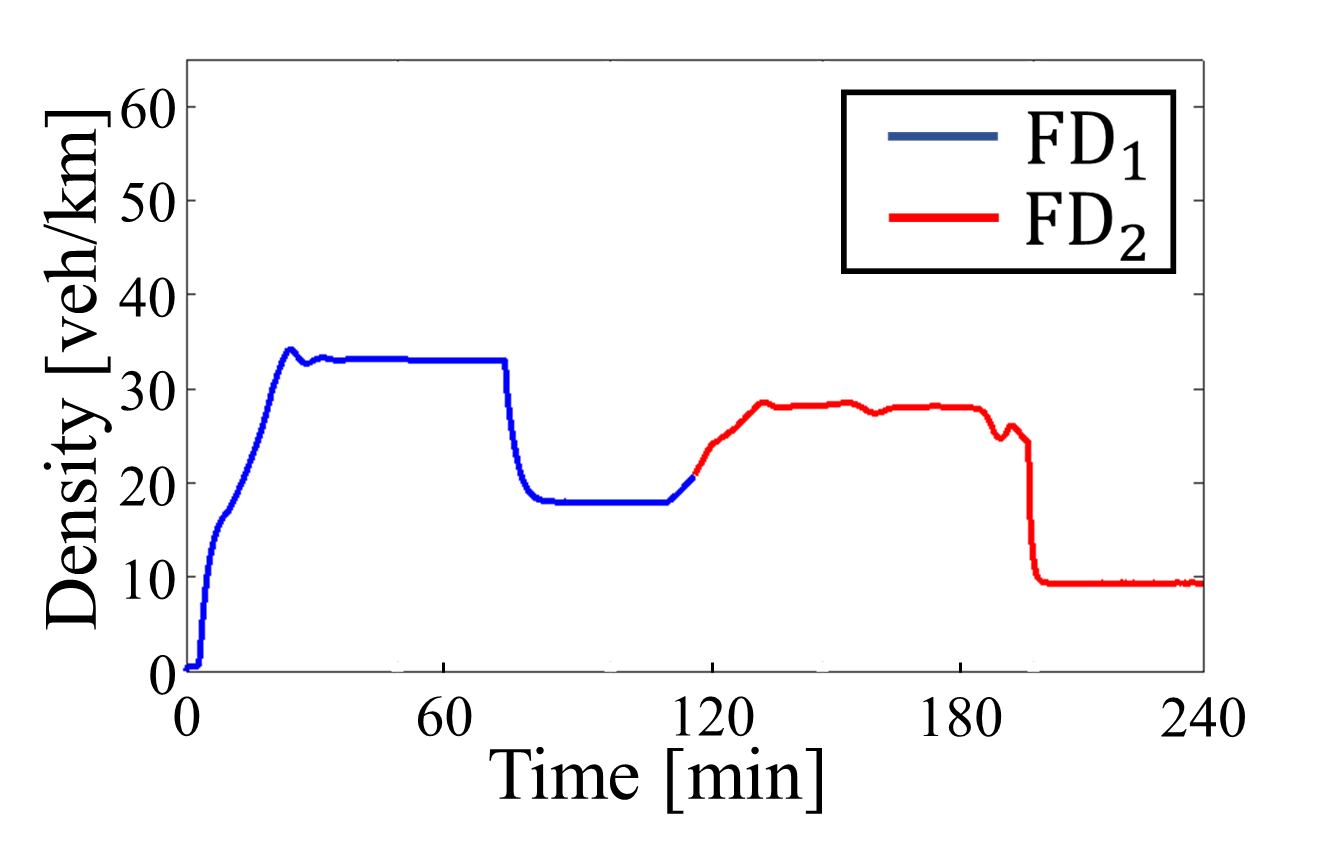}}\\
    
    \subfloat[]{\includegraphics[width =  0.49\linewidth]{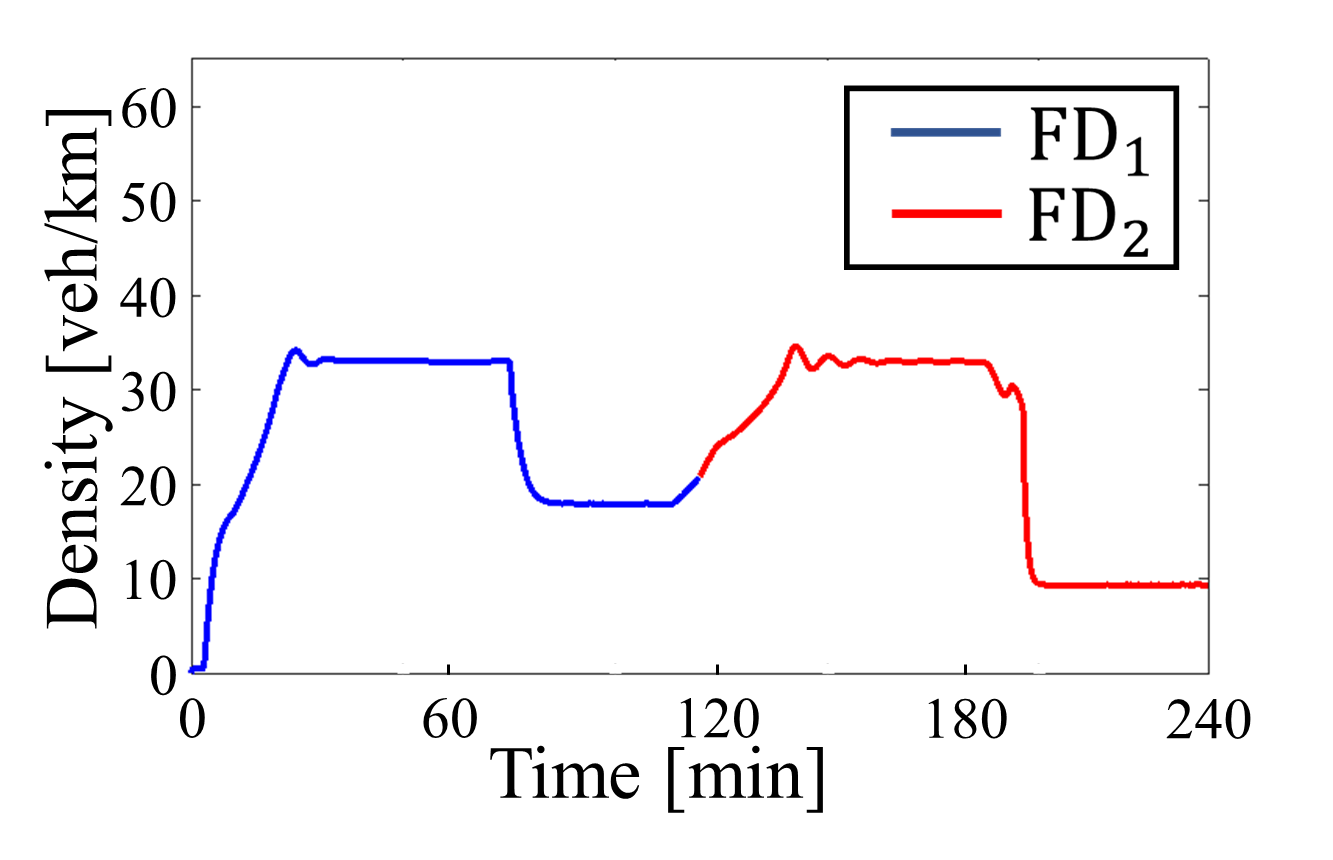}}
    \subfloat[]{\includegraphics[width =  0.49\linewidth]{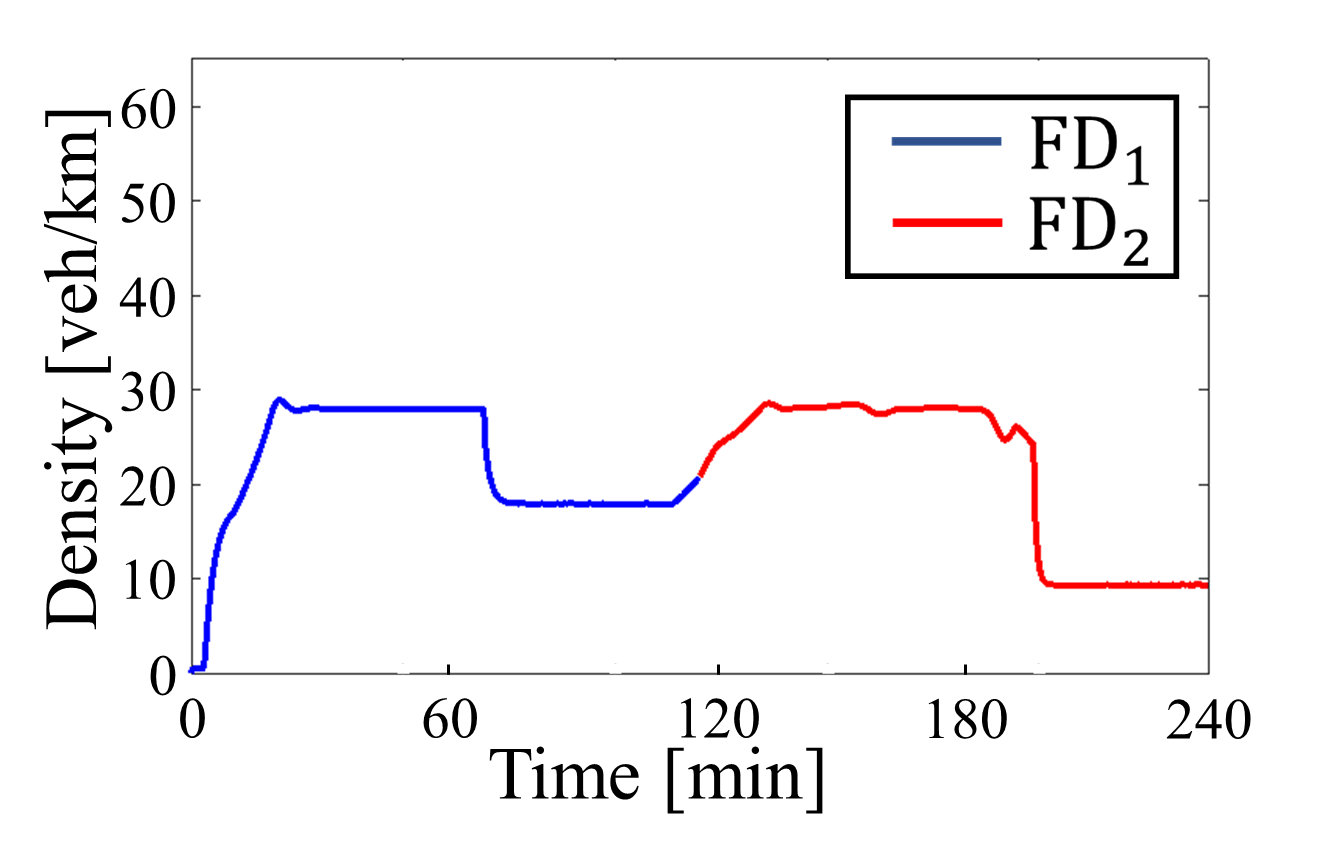}}\hspace{0 cm}\\
    
    \subfloat[]{\includegraphics[width =  0.49\linewidth]{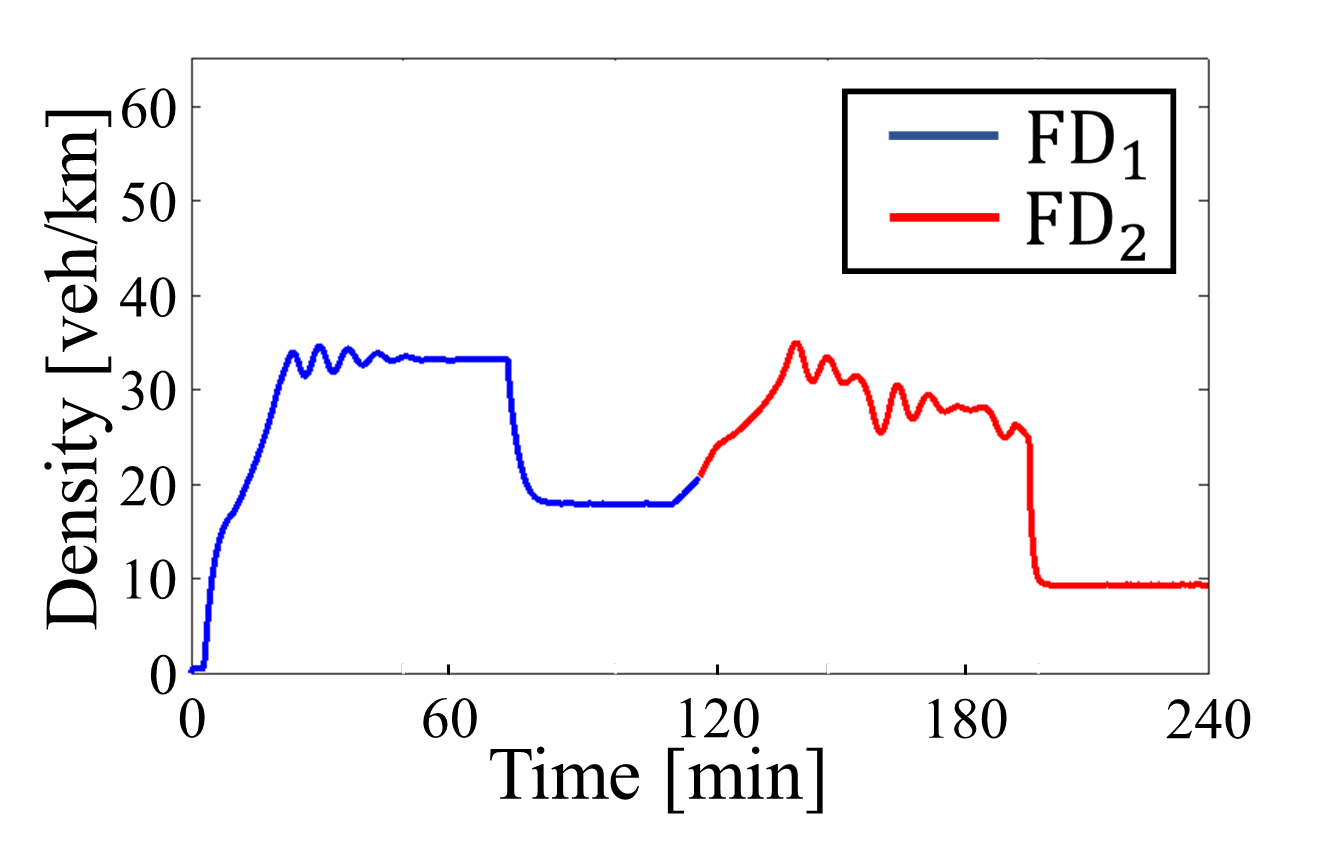}}
    \subfloat[]{\includegraphics[width =  0.49\linewidth]{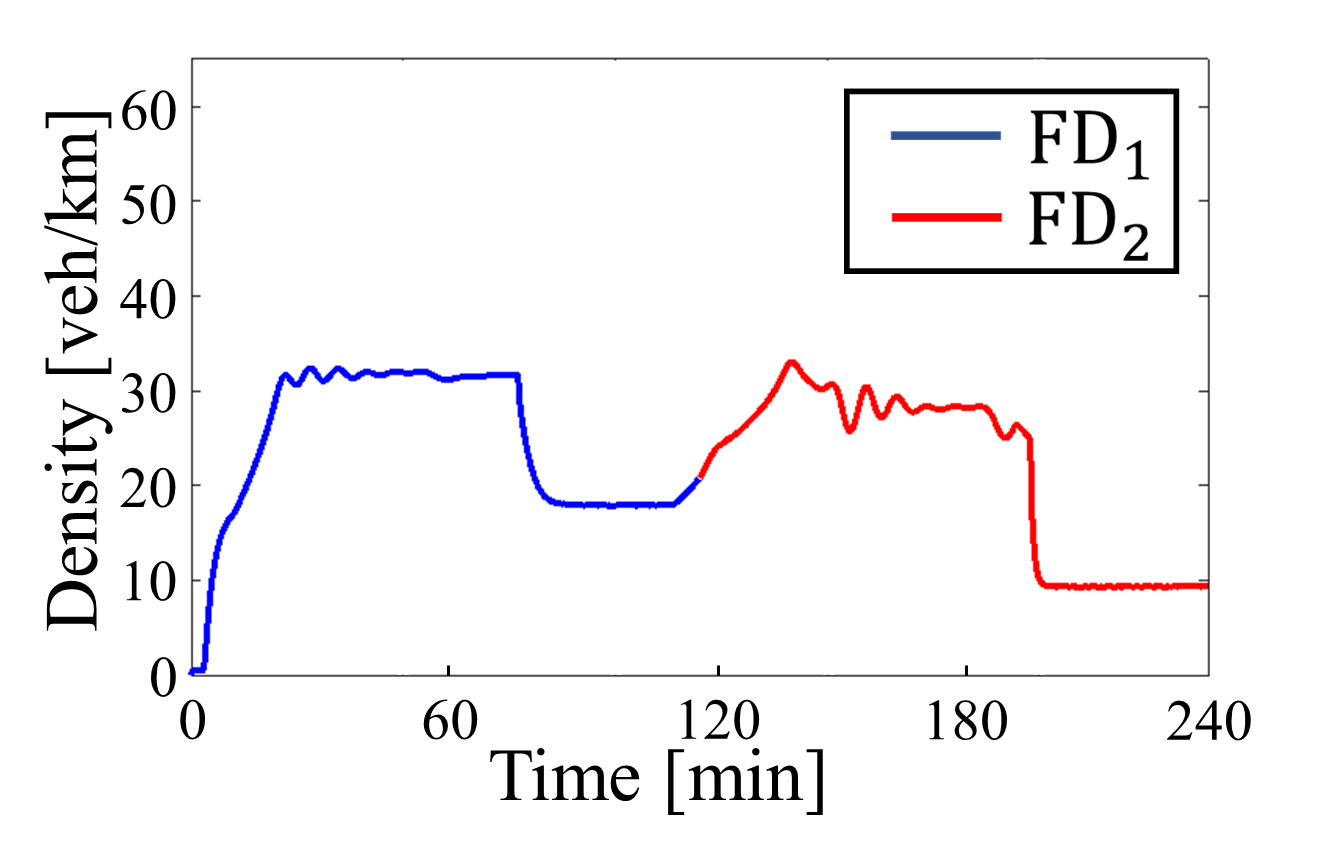}}\hspace{0 cm}
    \caption{Density of the bottleneck area: (a) Scenario~1. (b) Scenario~2. (c) Scenario~3-a. (d) Scenario~3-b. (e) Scenario~4-a. (f) Scenario~4-b.}
	\label{fig:DensityPerform}
\end{figure}

\begin{figure}[tb]
\centering
	 \subfloat[]{\includegraphics[width = 0.5\linewidth]{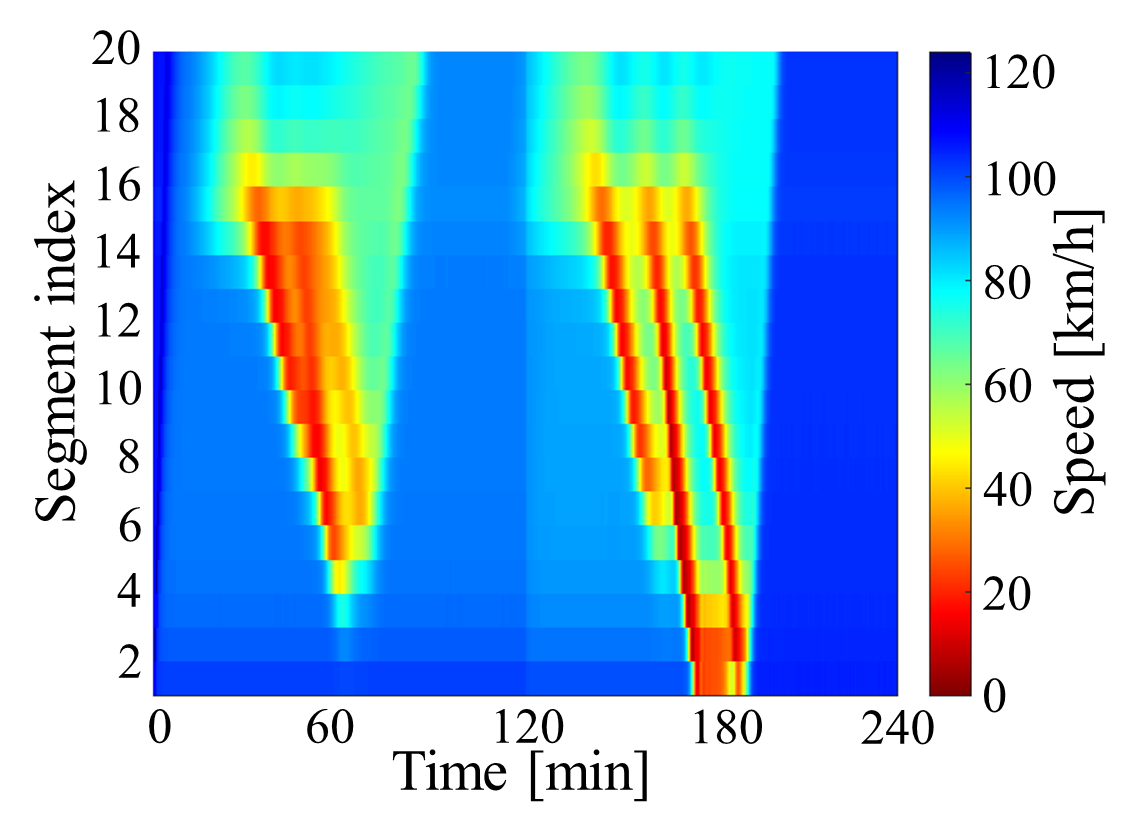}}
	 \subfloat[]{\includegraphics[width = 0.5\linewidth]{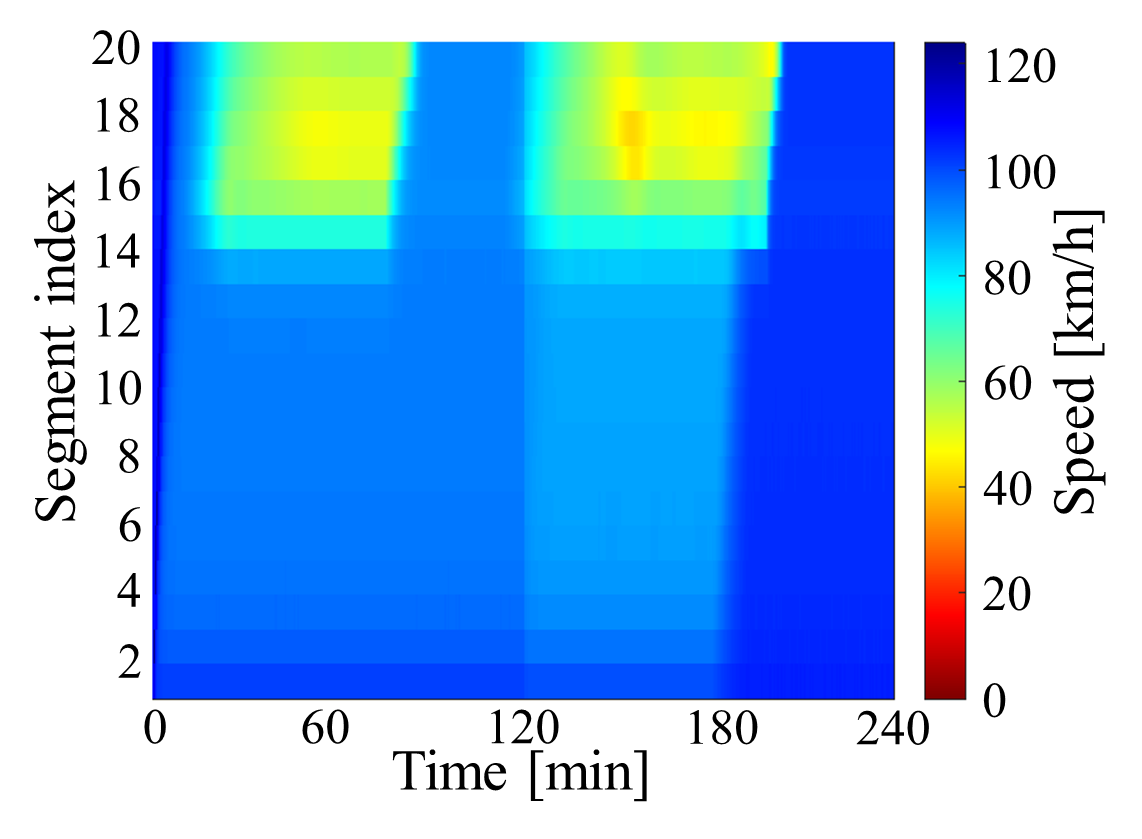}}\hspace{0 cm}\\
	 
    \subfloat[]{\includegraphics[width = 0.5\linewidth]{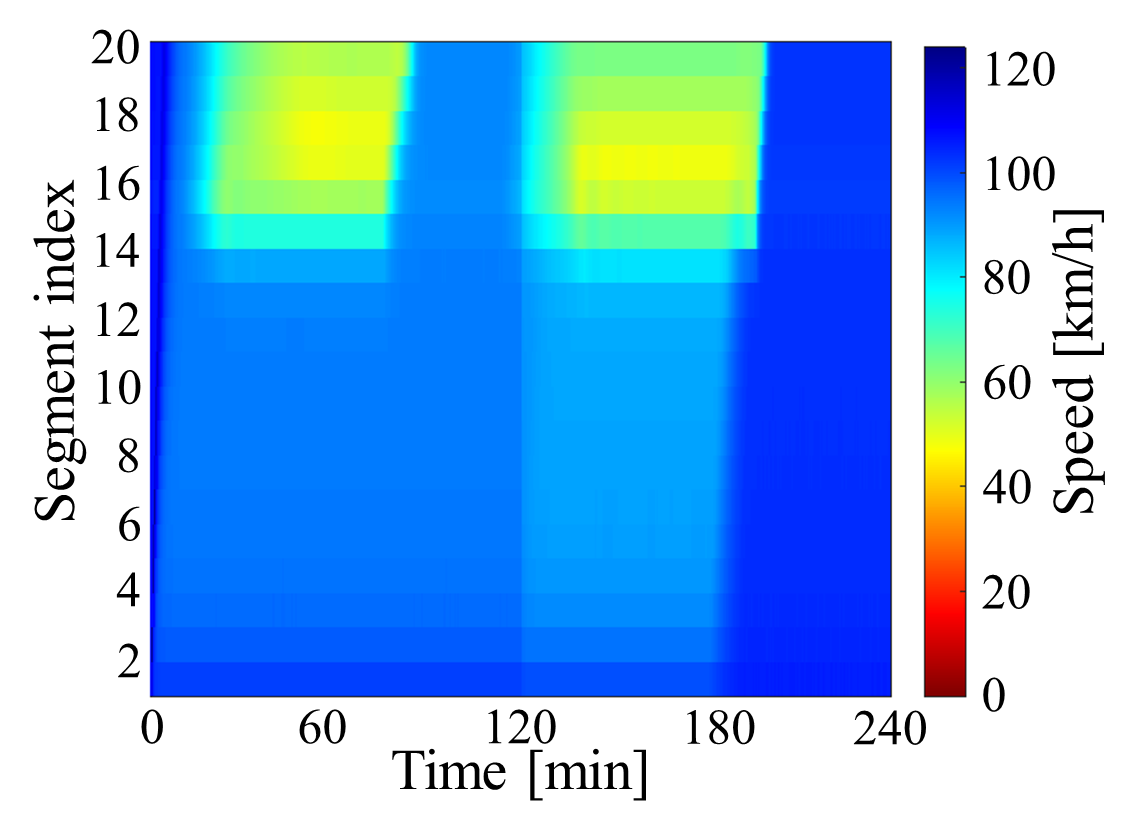}}
    \subfloat[]{\includegraphics[width = 0.5\linewidth]{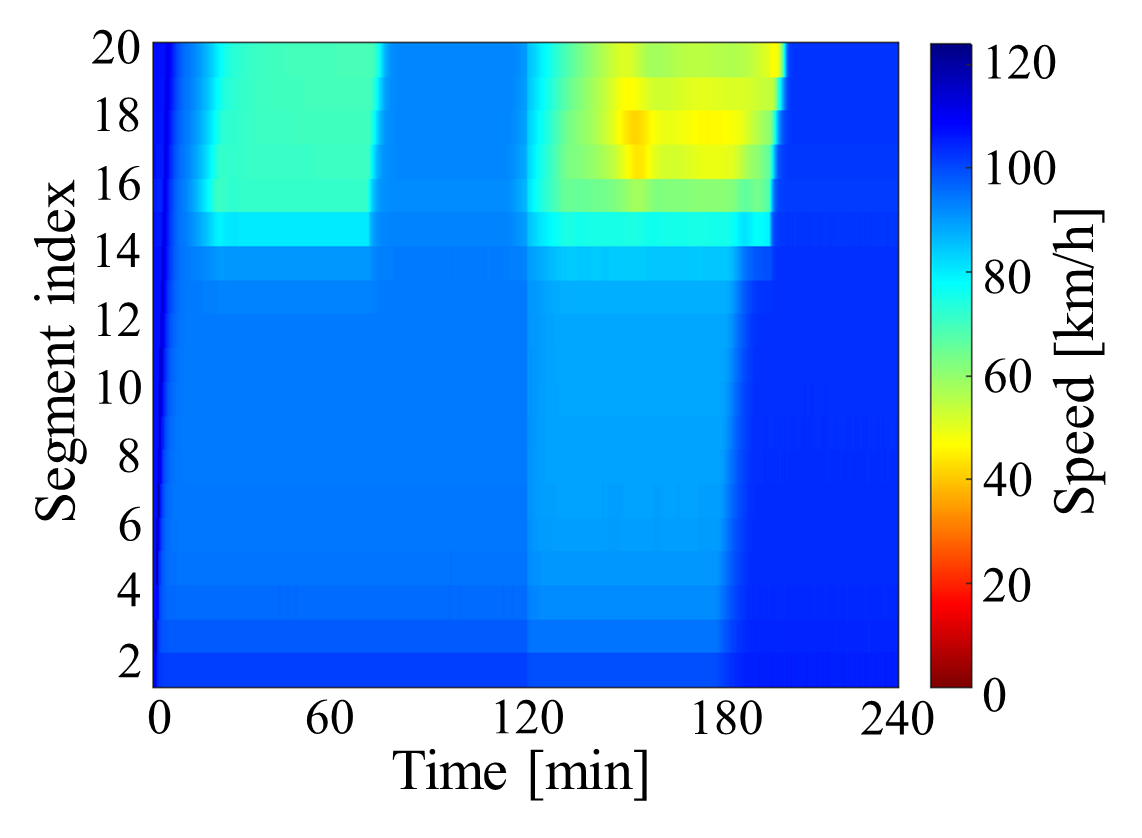}}\\
    
    \subfloat[]{\includegraphics[width = 0.5\linewidth]{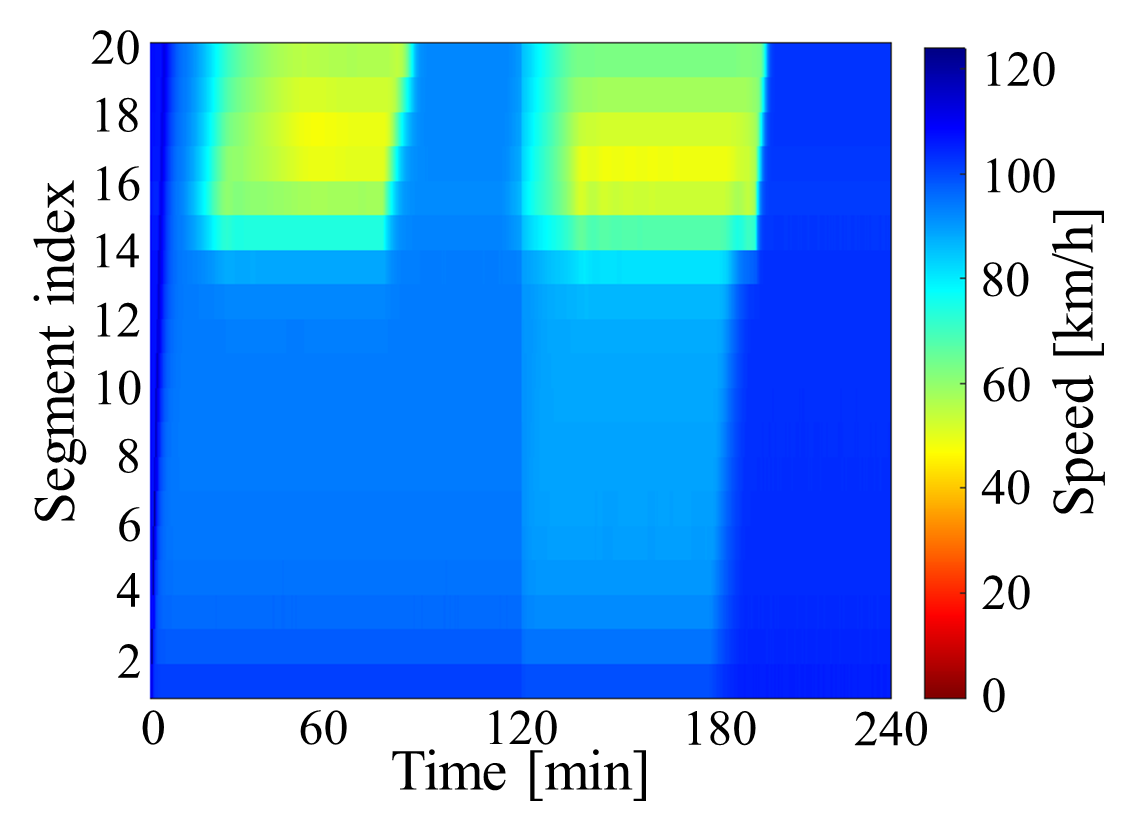}}
    \subfloat[]{\includegraphics[width = 0.5\linewidth]{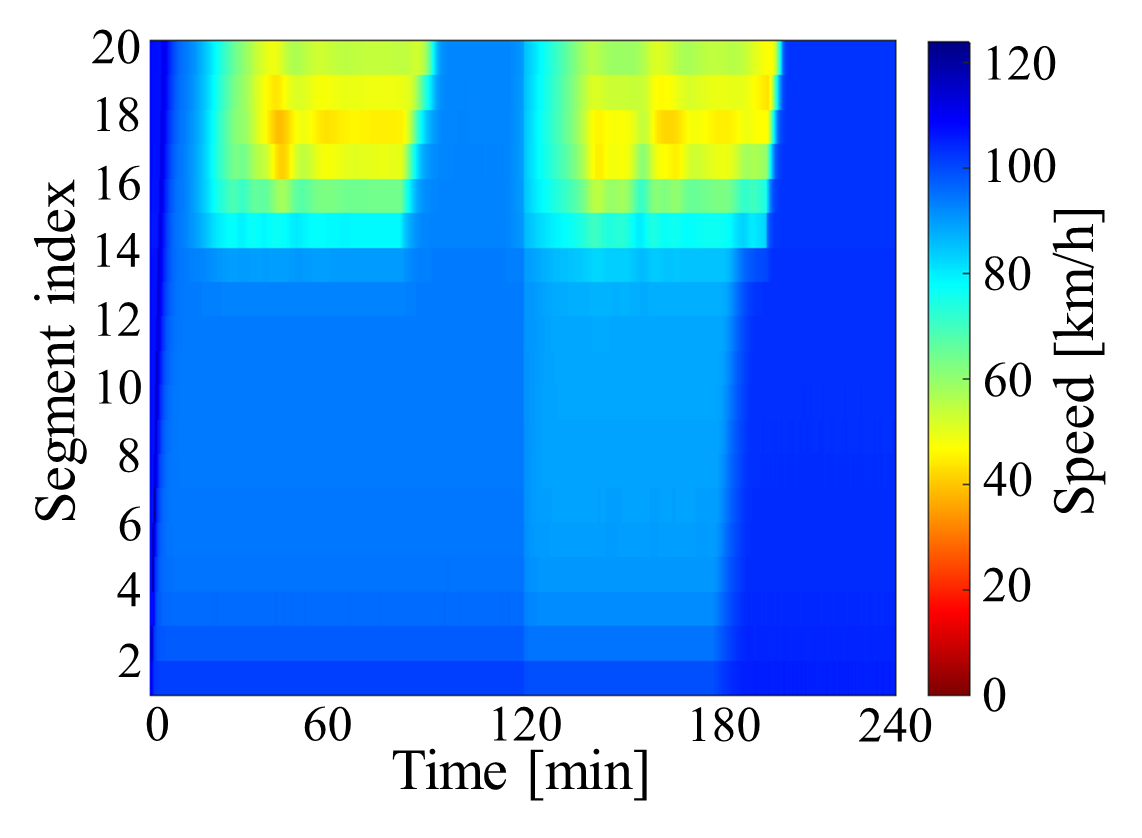}}
    \caption{Contour plots for speed: (a) Scenario~1. (b) Scenario~2. (c) Scenario~3-a. (d) Scenario~3-b. (e) Scenario~4-a. (f) Scenario~4-b.}
	\label{fig:contourPerform}
\end{figure}

\begin{figure}[tb]
\centering
	 \subfloat[]{\includegraphics[width = 0.49\linewidth]{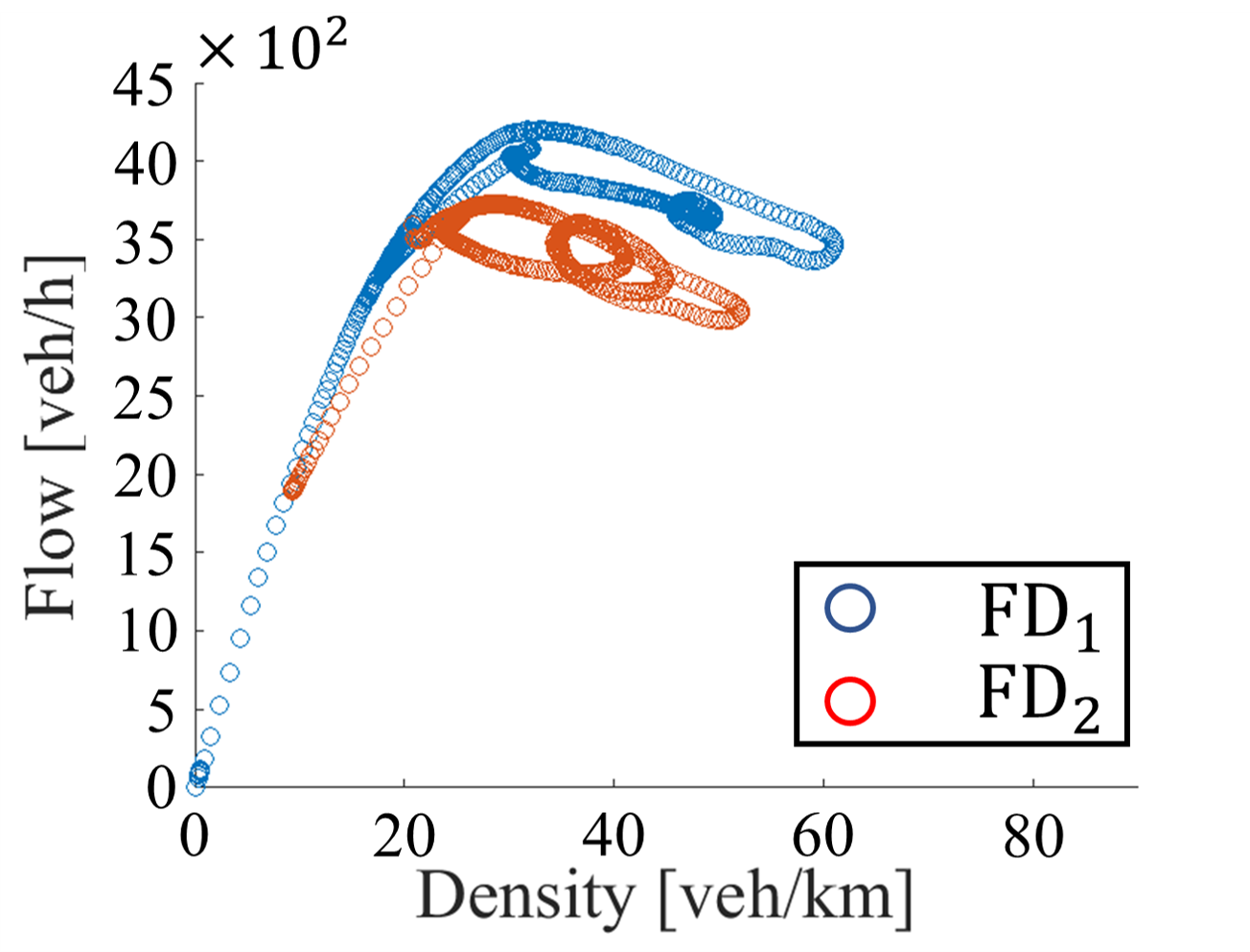}}
	 \subfloat[]{\includegraphics[width = 0.49\linewidth]{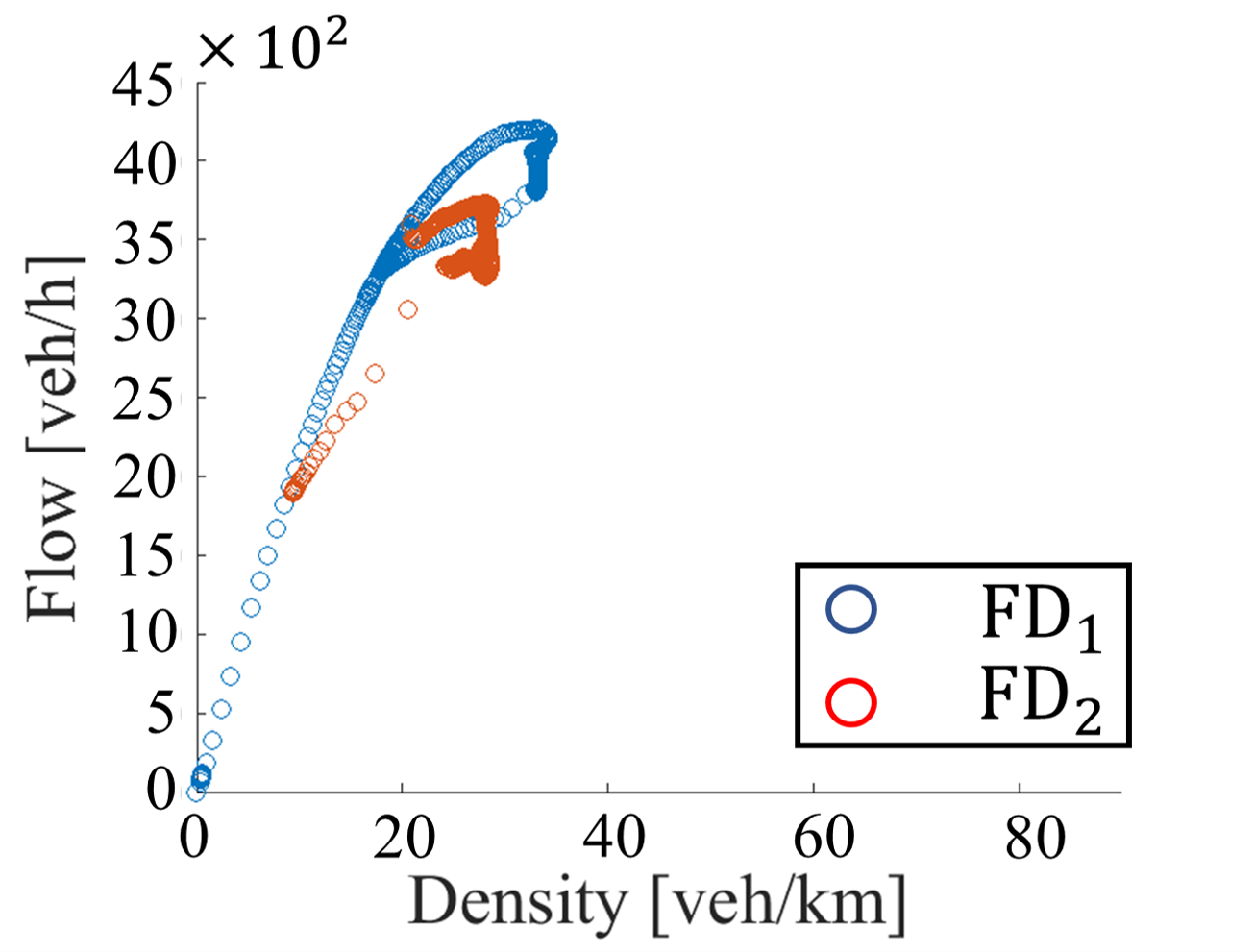}}\hspace{0 cm}\\
	 
    \subfloat[]{\includegraphics[width = 0.49\linewidth]{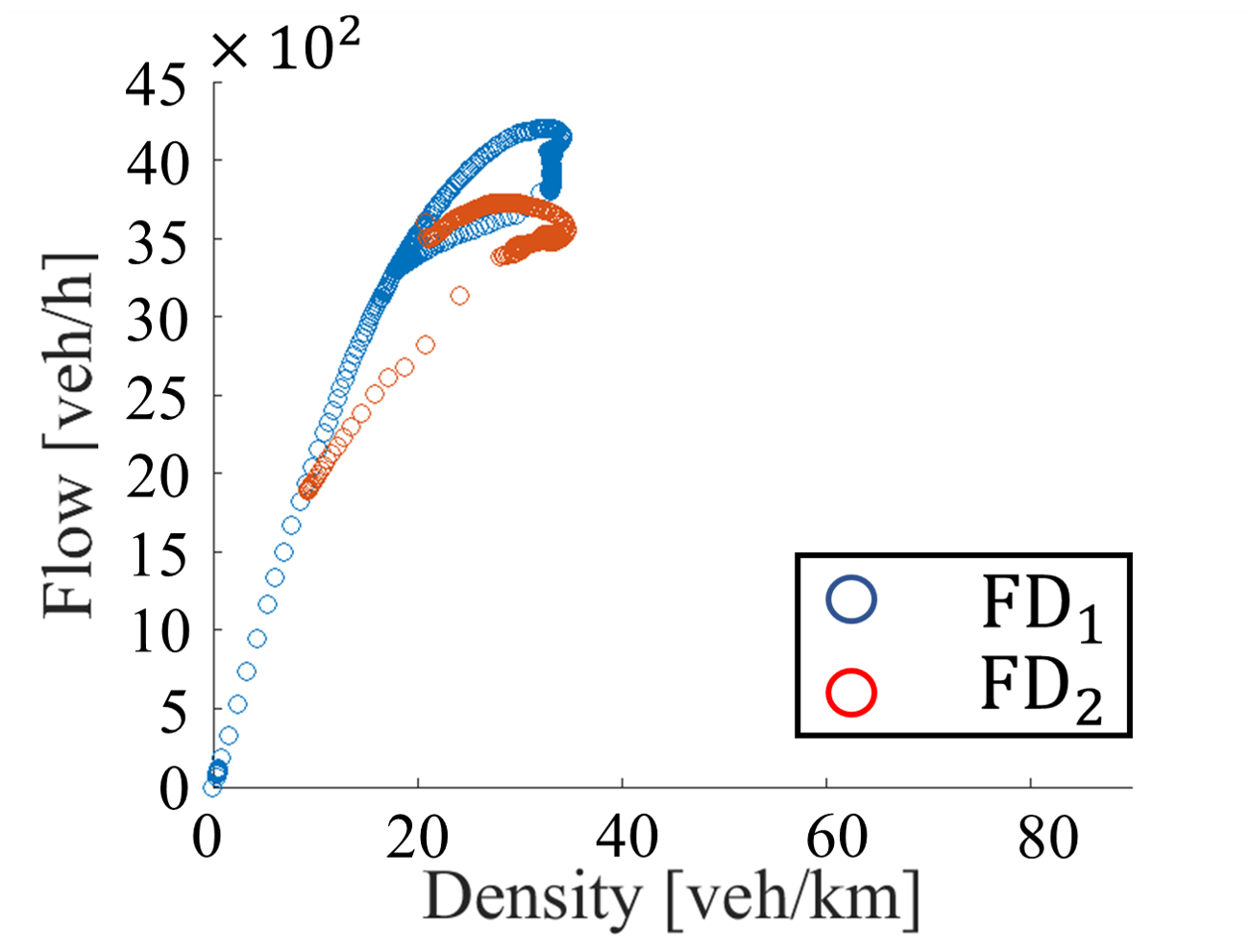}}\hspace{0 cm}
    \subfloat[]{\includegraphics[width = 0.49\linewidth]{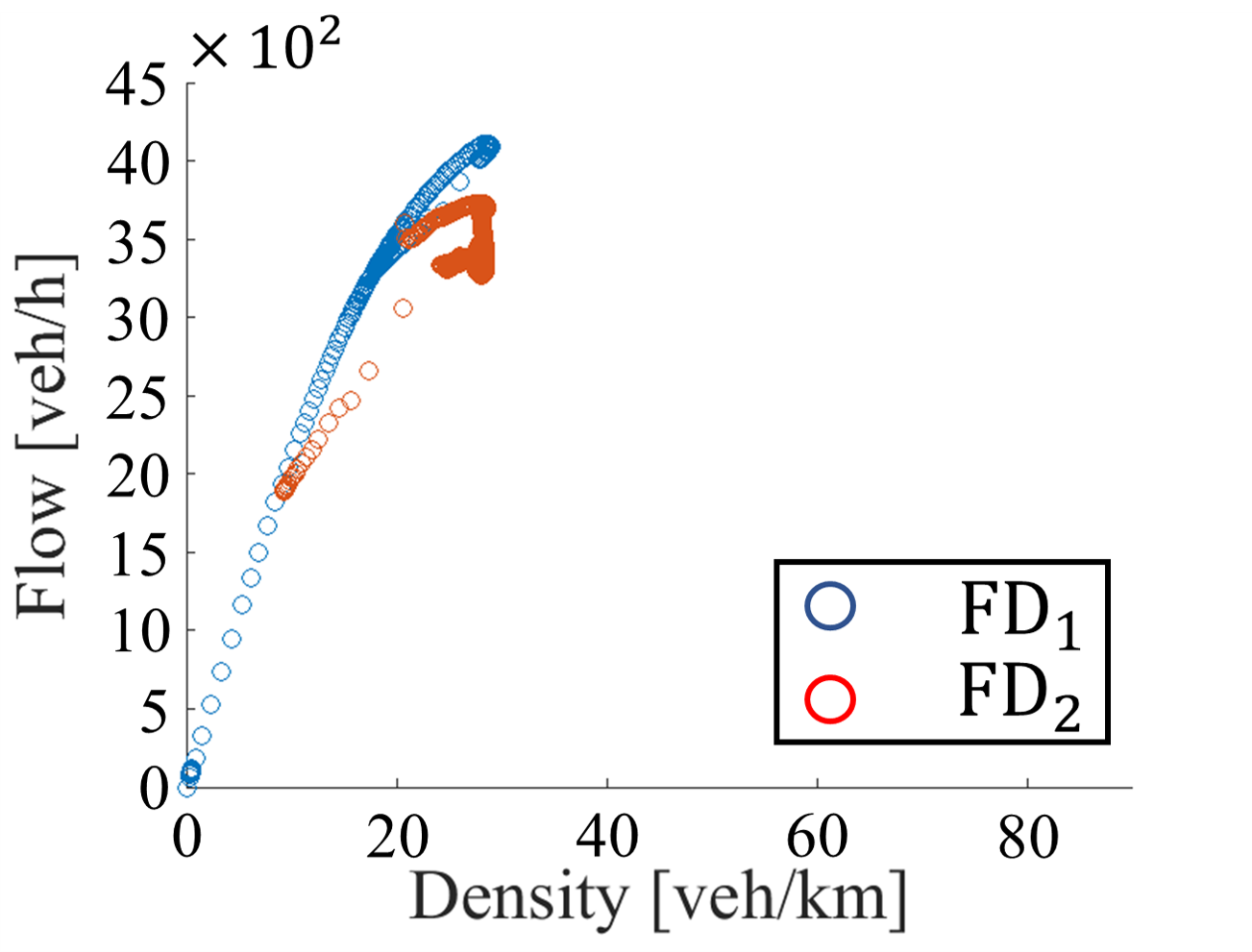}}\hspace{0 cm}\\
    
    \subfloat[]{\includegraphics[width = 0.49\linewidth]{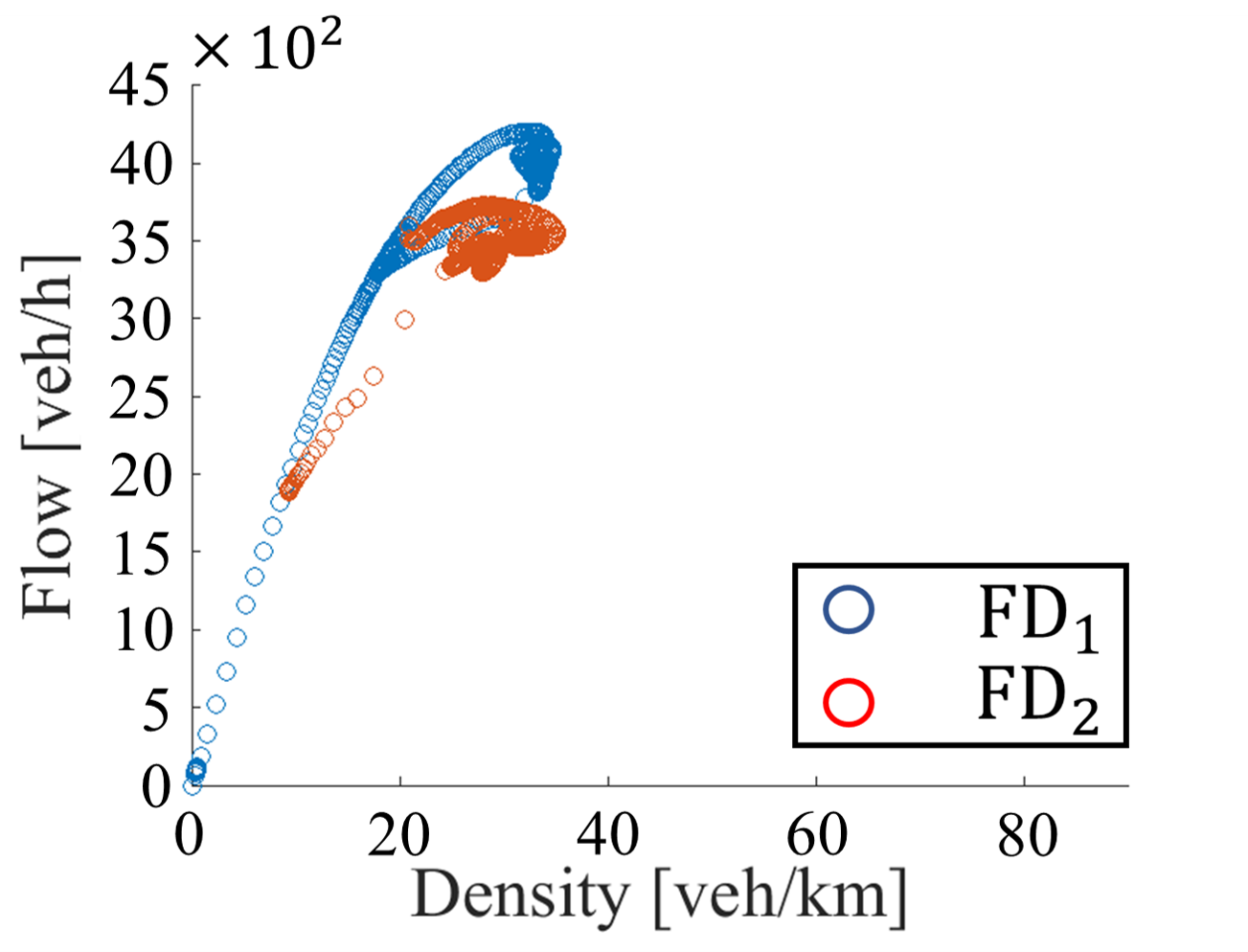}}\hspace{0 cm}
    \subfloat[]{\includegraphics[width = 0.49\linewidth]{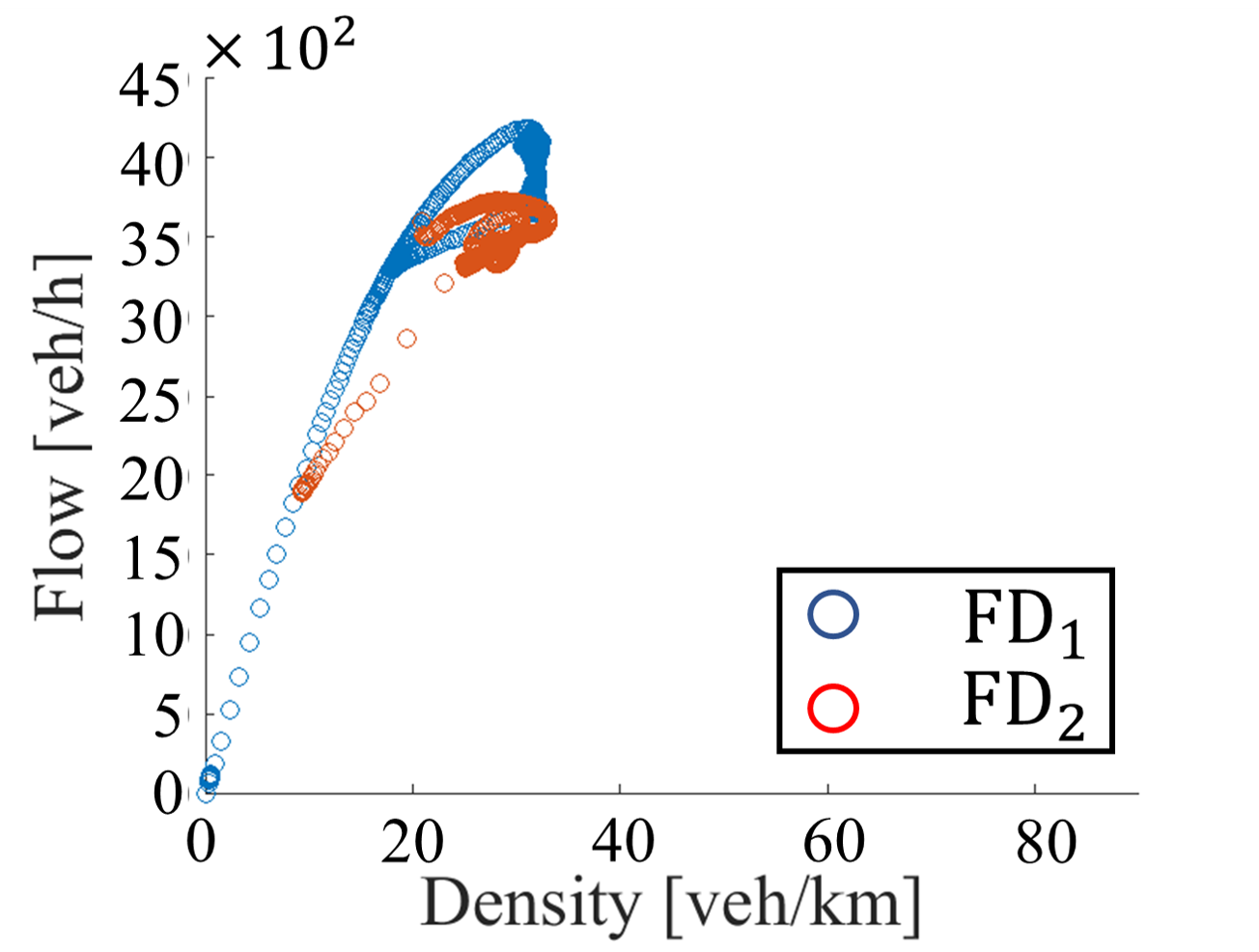}}\hspace{0 cm}
    
    \caption{Fundamental Diagram: (a) Scenario~1. (b) Scenario~2. (c) Scenario~3-a. (d) Scenario~3-b. (e) Scenario~4-a. (f) Scenario~4-b.}
	\label{fig:FD}
\end{figure}

The no-control case consists of the implementation of the nonlinear traffic model (\ref{eq:model1})-(\ref{eq:modelend}) in the presented motorway stretch, where no ramp metering is considered. According to Fig.~\ref{fig:contourPerform}(a), one may see that congestion occurs twice at the merging area (segment 15), which spills back upstream reaching up to segment~2, while the density at the bottleneck cell grows well above its critical value as it can be seen from Figs.~\ref{fig:DensityPerform}(a) and~\ref{fig:FD}(a).
The congestion occurs due to the high inflow entering both the mainstream and the ramp, which exceeds the bottleneck capacity. In fact, during the first half of the simulation, when the bottleneck capacity is around 4000~veh/h, the total demand reaches~4300~veh/h; in the second half of the simulation, when the bottleneck capacity is around 3600~veh/h, the total demand reaches~3800~veh/h. Note that capacity drop also happens at the bottleneck cells of the stretch, which reduces capacity once congestion is set, with the consequence of intensifying the resulting congestion.
The resulting TTS, calculated via~\eqref{eq:tts}, is reported in Table~\ref{tab:TTS}

\subsection{Scenario 2: Controlled cases with time-varying known set-points} \label{sec:Scenario2}

Analysing the results of Section~\ref{sec:no-control} and, in particular, by looking at Fig.~\ref{fig:FD}(a), which shows FD$_1$ (blue) and FD$_2$ (red) resulting from the no-control case, one may observe that the actual critical densities of the FDs, i.e., the densities corresponding to the maximum outflows are 33~veh/km and 28~veh/km for FD$_1$ and FD$_2$, respectively. These values are the ones employed for the controlled case with known set-points.

Then, we evaluate the performance achievable by controlling the traffic via ramp metering, assuming that we have perfect knowledge of the critical densities (thus, the set-points) in real-time. Note that this corresponds to an unrealistic scenario, as the actual critical densities cannot be observed unless we are reaching a (nearly) congested state. Still, it is interesting to perform such an experiment, in order to determine an upper bound for the performance of our estimation strategy.

We therefore implement the nonlinear traffic model (\ref{eq:model1})~-~(\ref{eq:modelend}), where the on-ramp flow is calculated via \eqref{eq:alinea}, while $\hat{\rho}^{\star}(k) = 33$~veh/km for $0 \leq k < 720 $ and $\hat{\rho}^{\star}(k) = 28$~veh/km for $720 \leq k \leq 1440$.

The results in Fig.~\ref{fig:DensityPerform}(b), Figs.~\ref{fig:contourPerform}(b), and Figs.~\ref{fig:FD}(b), show that congestion disappears and the bottleneck cell's density is maintained around its critical value for both FD$_1$ and FD$_2$. 
To assess the controller performance numerically, we compare the TTS, reported in Table~\ref{tab:TTS}, where one may see that Scenario~2 results in a 5.9\% improvement over the no-control case in Scenario~1.
Furthermore, queues are generated at the on-ramp location during the peak periods in all controlled scenarios; note that no upper bound for the queue length are considered in our experiments. 

\subsection{Scenario 3: Controlled with constant set-points} \label{sec:Scenario3}

In this scenario, we apply ramp metering employing a constant set-point during the simulation. Basically, this scenario represents what is typically done in existing ramp metering applications, where the set-point is estimated from historical data and maintained constant during implementation.
In particular, we test two sub-scenarios, one (Scenario~3-a) using as a set-point the critical density of FD$_1$ and another one (Scenario~3-b) using as a set-point the critical density of FD$_2$.
That is, we implement the nonlinear traffic model (\ref{eq:model1})-(\ref{eq:modelend}), where the on-ramp flow is calculated via \eqref{eq:alinea}, where, in Scenario~3-a, $\hat{\rho}^{\star}(k) = 33$~veh/km, $\forall k $; and, in Scenario~3-b, $\hat{\rho}^{\star}(k) = 28$~veh/km, $\forall k $.

Results in Fig.~\ref{fig:DensityPerform}(c,d) show that the congestion is mitigated in both sub-scenarios. According to Figs.~\ref{fig:contourPerform}(c,d) and Figs.~\ref{fig:FD}(c,d), we observe that, for each sub-scenarios, the controller is capable to maintain the bottleneck cell's density around the desired set-point during the period characterised by high demand for both FD$_1$ to FD$_2$.
However, as these values do not maximise the throughput for some time, numerical comparisons presented in~Table~\ref{tab:TTS} reveal that the controller is capable to achieve only a 1.9\% and 2.5\% TTS improvement compared to the no-control case for Scenarios~3-a and~3-b, respectively.  

\subsection{Scenario 4: Controlled with estimated set-points} \label{sec:Scenario4}

We proceed then with evaluating the performance of our estimator, by considering two sub-scenarios considering different initial values for the estimated set-points, corresponding to the critical densities of FD$_1$ and FD$_2$.
We implement the nonlinear traffic model (\ref{eq:model1})-(\ref{eq:modelend}), where the on-ramp flow is calculated via \eqref{eq:alinea}, and $\hat{\rho}^{\star}(k)$ and $\hat{\rho}^{\star}(k)$ are estimated via \eqref{eq:rhoStar} and \eqref{eq:qStar} respectively.
We test Scenario~4-a, where $\hat{\rho}^{\star}(0)=33$~veh/km, and Scenario~4-b, where $\hat{\rho}^{\star}(0)=28$~veh/km.

\begin{figure}[tb]
		\centering
    \subfloat[]{\includegraphics[width = 0.49\linewidth]{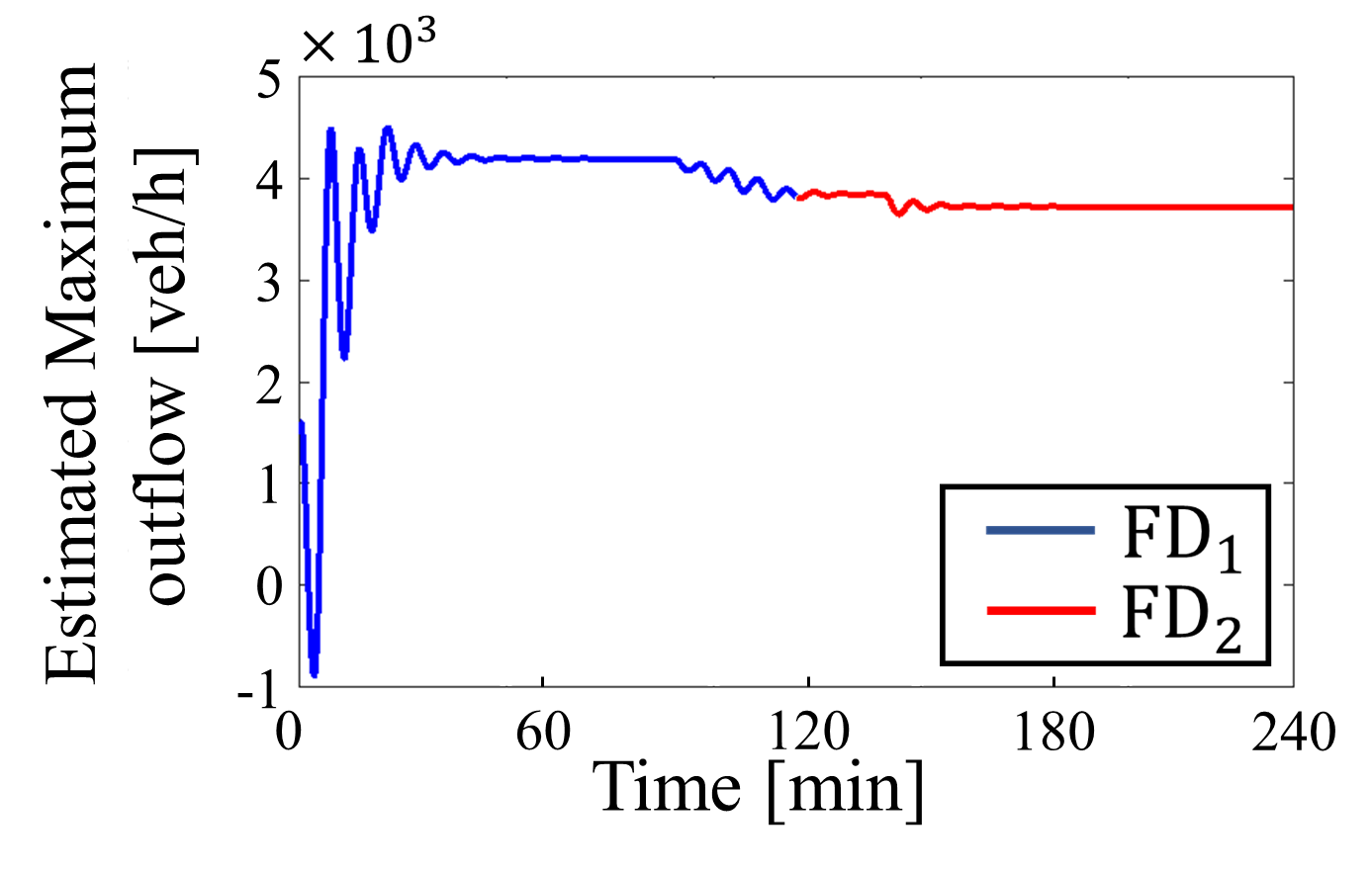}}\hspace{0cm}
    \subfloat[]{\includegraphics[width = 0.49\linewidth]{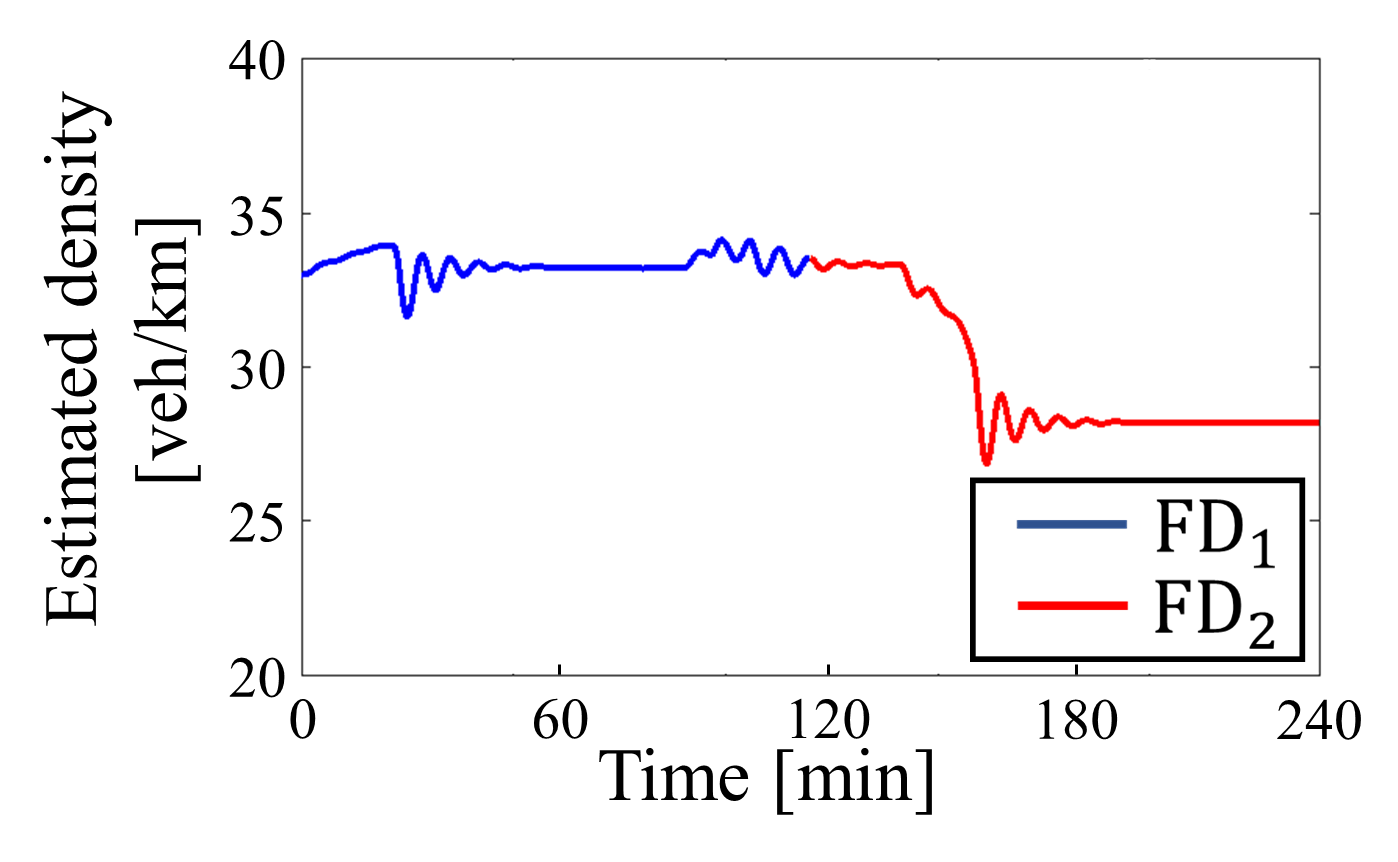}}\hspace{0cm} \\
    \subfloat[]{\includegraphics[width = 0.49\linewidth]{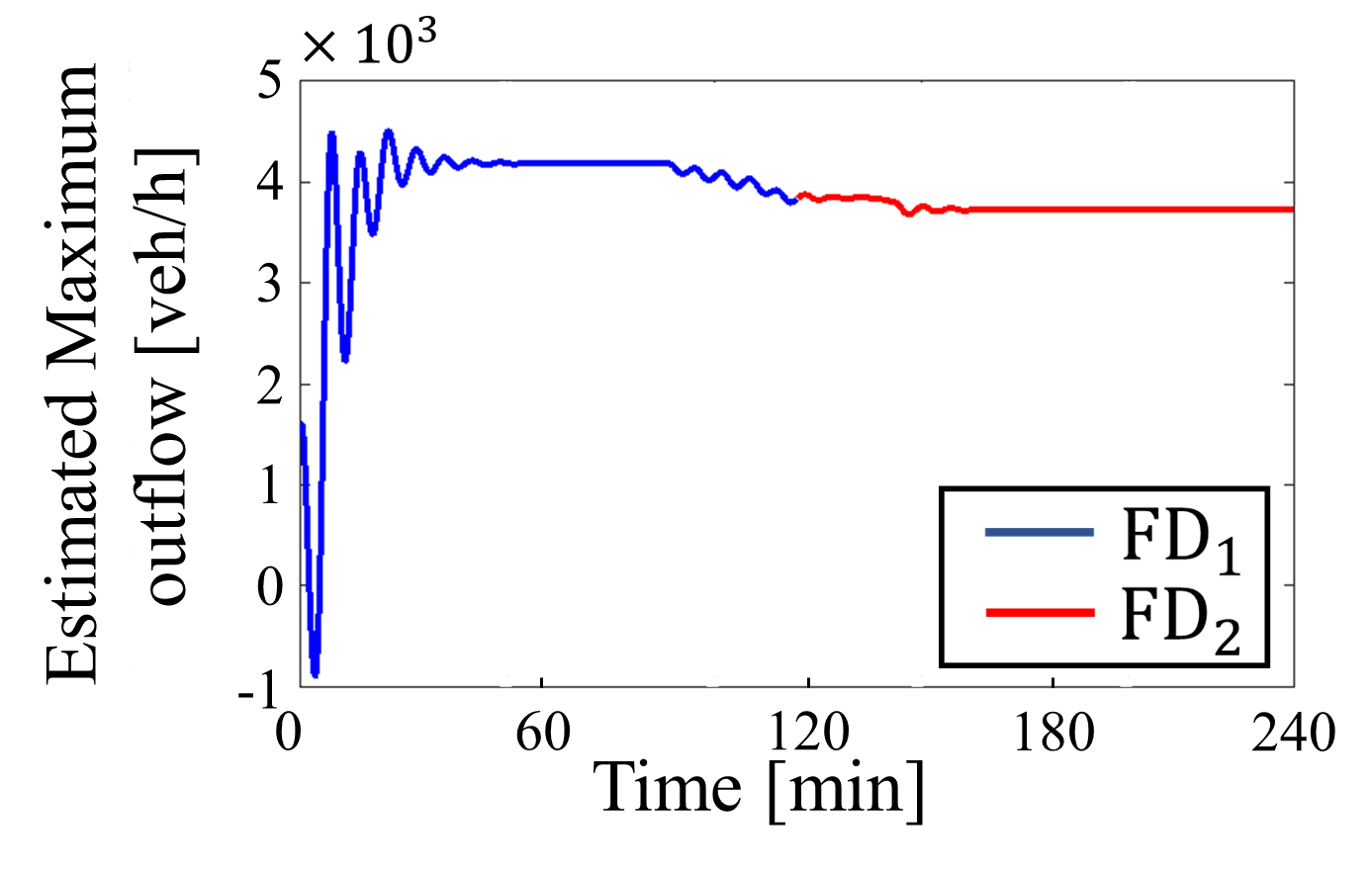}}\hspace{0cm}
    \subfloat[]{\includegraphics[width = 0.49\linewidth]{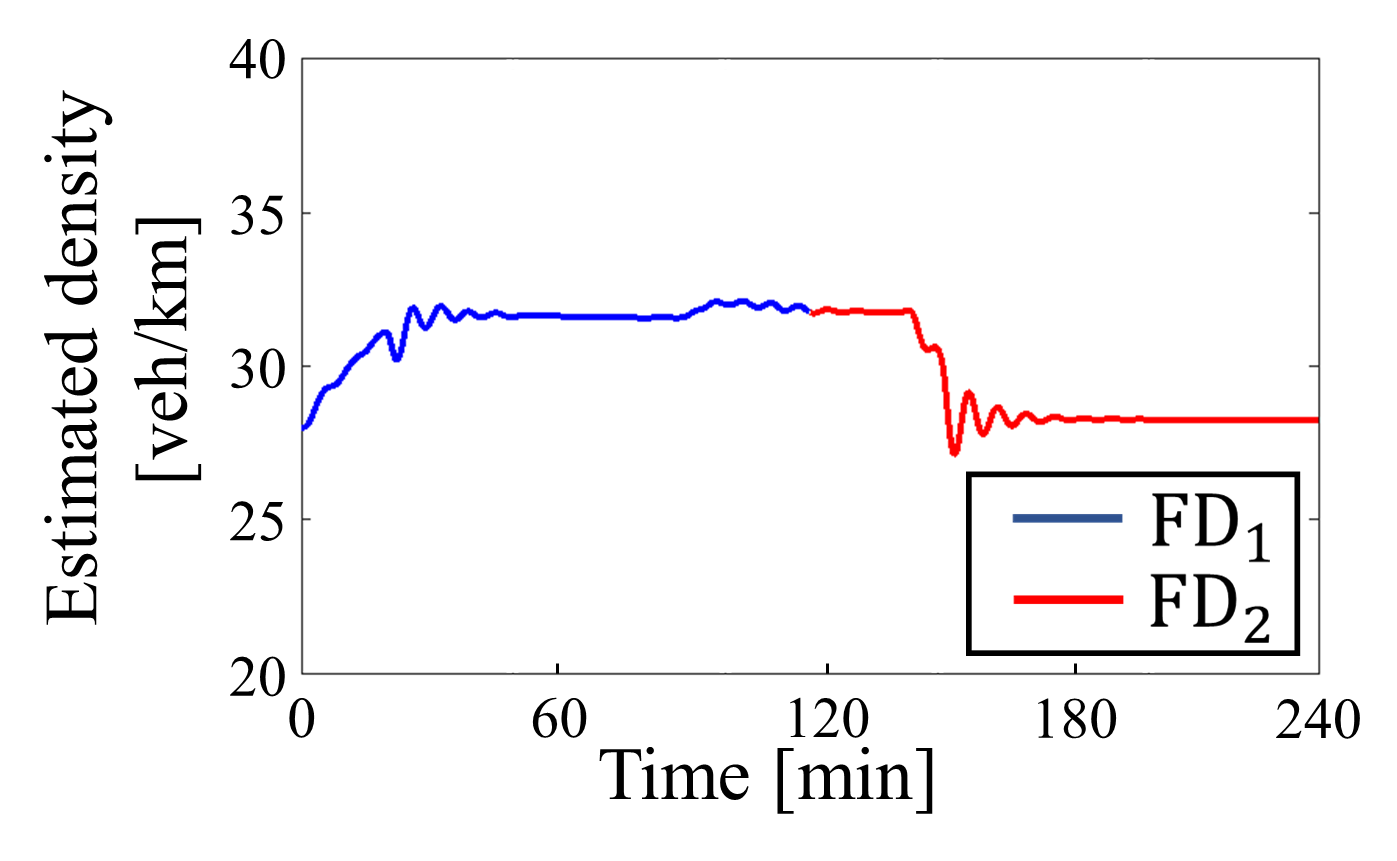}}\hspace{0cm} \\
	\caption{Estimated values, left: maximum out flow ($q^{\star}$), right: Critical density ($\rho^{\star}$), for: (a,b) Scenario~4-a, where $\hat{\rho}^{\star}(0) = 33$ veh/km. (c,d) Scenario~4-b, where $\hat{\rho}^{\star}(0) = 28$ veh/km.}
	\label{fig:device}
\end{figure}

Looking at the results regarding Scenario~4-a, as shown in Figs.~\ref{fig:DensityPerform}(e),~\ref{fig:contourPerform}(e), and~\ref{fig:FD}(e), we observe that the controller with the estimator is capable of avoiding the onset of congestion, similarly to the other controlled scenarios. 
Moreover, we can also see that the estimator manages to successfully adjust set-point values to the actual critical values, while successfully controlling the system. 
This is also shown in more detail in Fig.~\ref{fig:device}(b), where the estimated critical density settles to the actual value around $t = 150$ min, i.e., 30 minutes after the change in the FD. In addition, Fig.~\ref{fig:device}(a) shows the estimated maximum outflow, where the changes start earlier than the estimated density ($t = 120$ min). 
According to the figure, the estimated maximum flow at the beginning of the simulation reaches some negative values; however, note that this has no impact on the performance of the controller since the estimated critical density, which is used as a set-point, assumes  always positive values. 
Numerical comparisons in terms of TTS, reported in Table~\ref{tab:TTS}, demonstrate that utilising the estimator not only improves traffic conditions compared to the control case (Scenario 1), but also outperforms all the scenarios where a constant set-point is used; for example, the TTS improvement in Scenario~4-a is 63\% higher than in Scenario~3-a.

	
    
\begin{table}[tb]
	\renewcommand{\arraystretch}{1.3}
	\centering
	\caption{TTS value report regarding the different scenarios.}
	\label{tab:TTS}

\input{TTS}
\end{table}


Similarly, in Scenario~4-b the controller manages to avoid the congestion successfully, as can be seen from Figs.~\ref{fig:DensityPerform}(f),~\ref{fig:contourPerform}(f), and~\ref{fig:FD}(f). 
In particular, as shown in Fig.~\ref{fig:DensityPerform}(f), the estimator successfully adjusts set-point values to the critical density after the FD changes. This is shown in more detail in Fig.~\ref{fig:device}(d), where the estimated critical density reaches first, at $t=20$ min, 33 veh/km and then, around $t = 140$ min, 28 veh/km. In addition, Fig.~\ref{fig:device}(c) reveals the estimated maximum outflow, where the convergence to its true value starts earlier than the estimated density ($t = 120$ min). 
Also for this Scenario, the resulting TTS is lower than the no-control case and than any controlled Scenario with a constant set-point; in particular, the TTS improvement in Scenario~4-b is 116\% higher than in Scenario~3-b.

\subsection{Scenario 5: Controlled with estimated set-points and distant initial values}

In this scenario, we investigate the performance of the controller with the adaptive estimator in two cases, where the initial values of the set-points are distant from the actual values, utilising in particular (a) a very high value (Scenario~5-a), i.e., $\hat{\rho}^{\star}(0) = 40$ veh/km, and (b) a very low value (Scenario~5-b), i.e., $\hat{\rho}^{\star}(0) = 20$ veh/km.

\begin{figure}[tb]
		\centering
    \subfloat[]{\includegraphics[width = 0.49\linewidth]{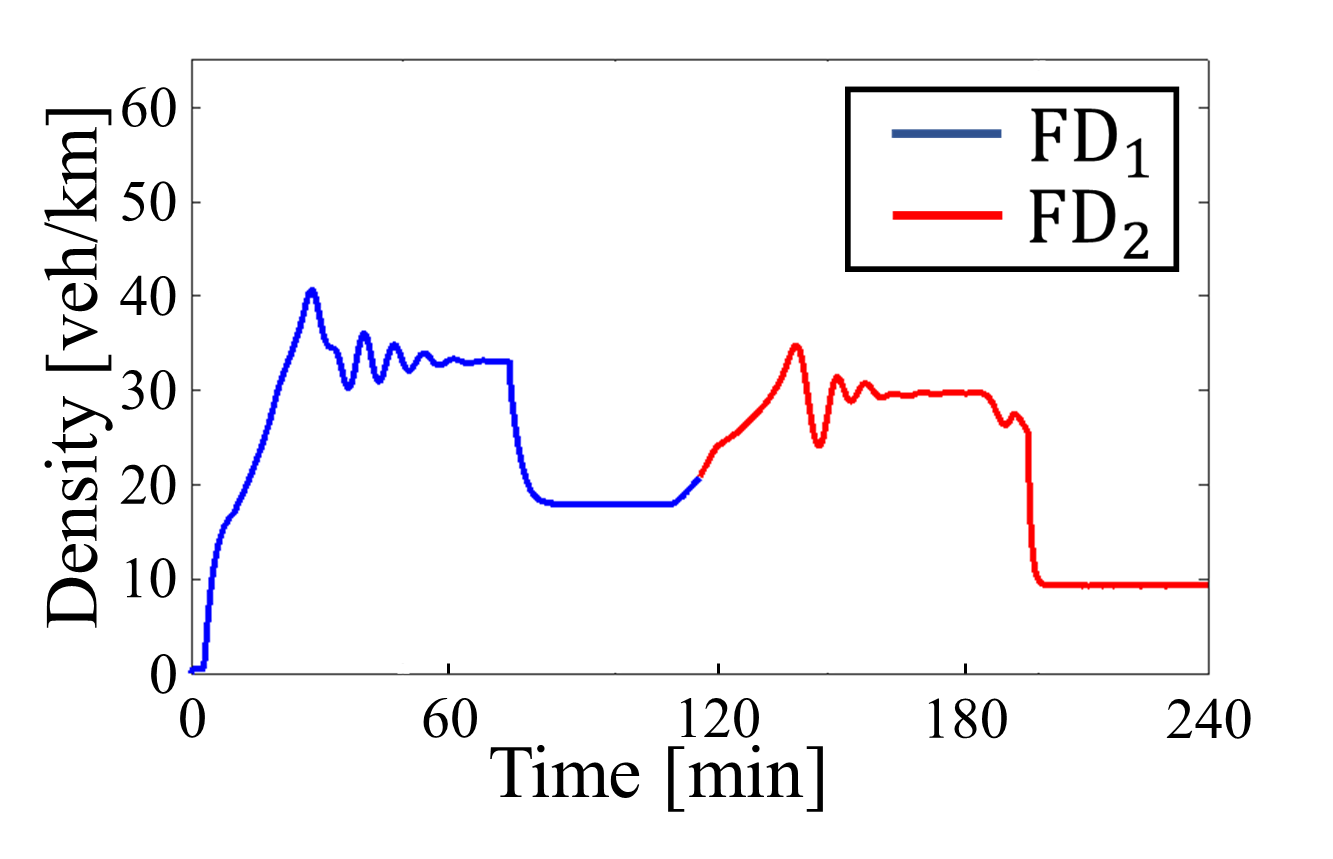}}\hspace{0 cm} 
    \subfloat[]{\includegraphics[width =  0.49\linewidth]{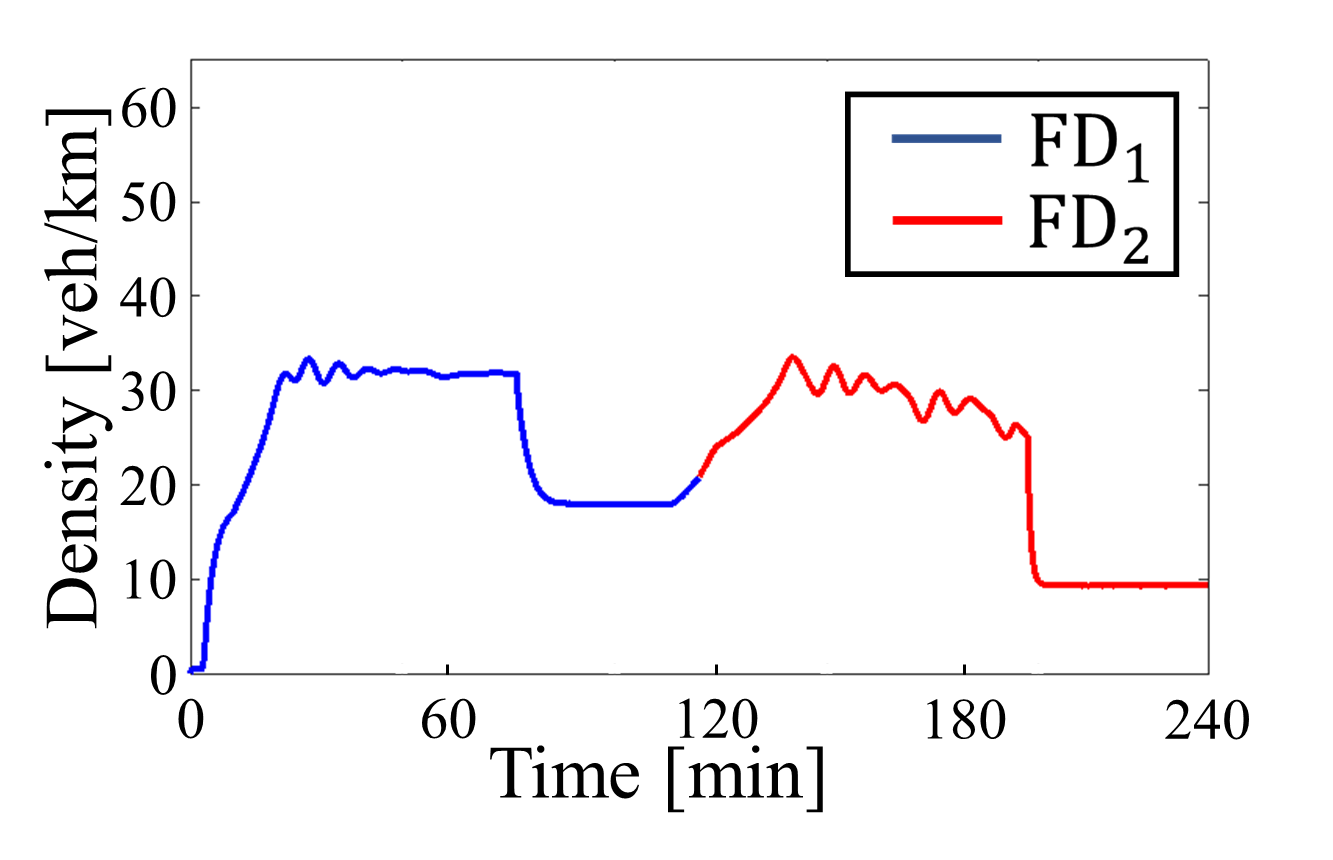}}\hspace{0 cm}
	\caption{Density of the bottleneck areas for: (a) Scenario 5-a, where $\hat{\rho}^{\star}(0) = 40$ veh/km. (b) Scenario 5-b, where $\hat{\rho}^{\star}(0) = 20$ veh/km.}
	\label{fig:DensityPerformExtreme}
\end{figure}

\begin{figure}[tb]
		\centering
    \subfloat[]{\includegraphics[width = 0.49\linewidth]{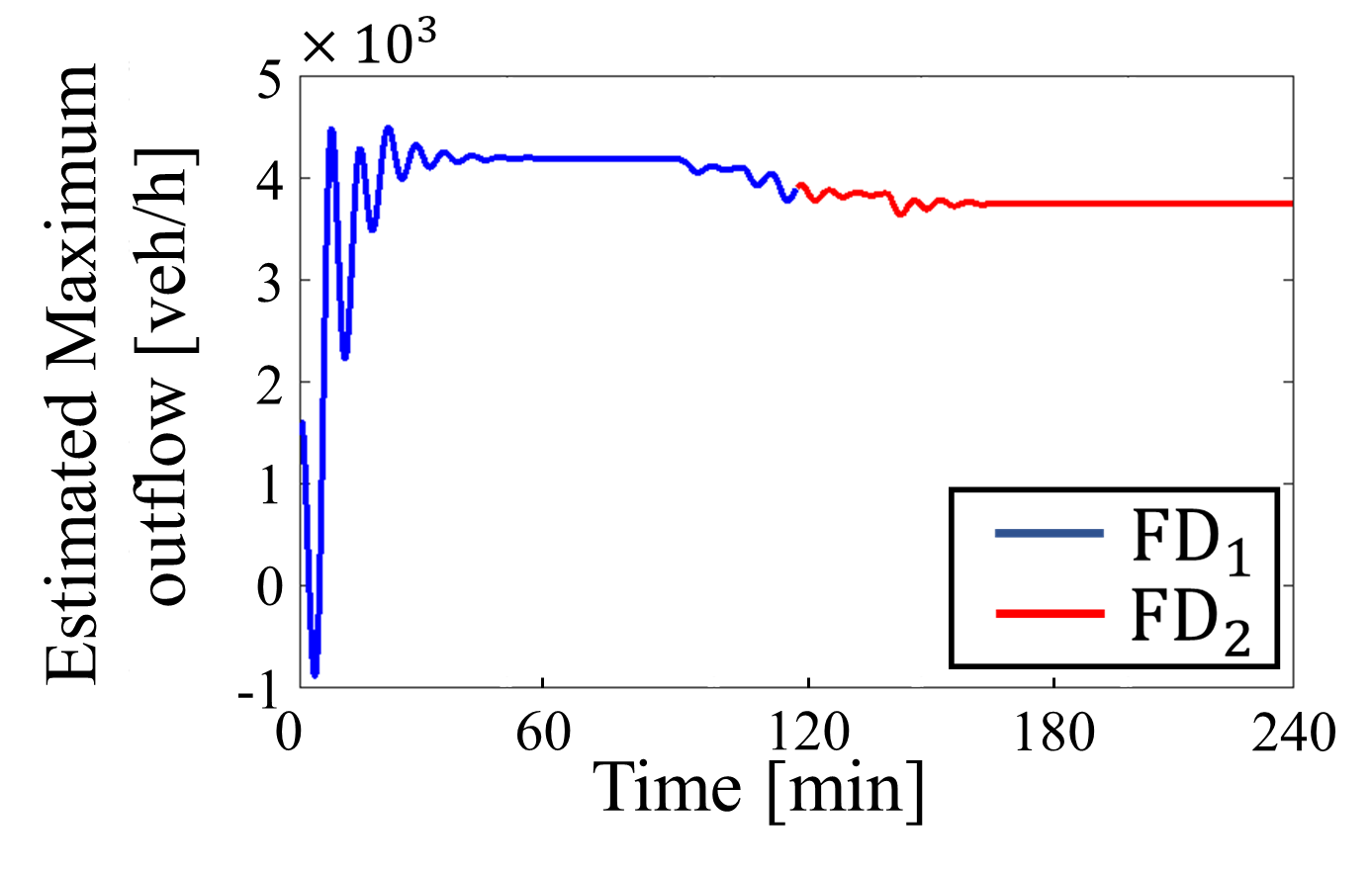}}\hspace{0cm}
    \subfloat[]{\includegraphics[width = 0.49\linewidth]{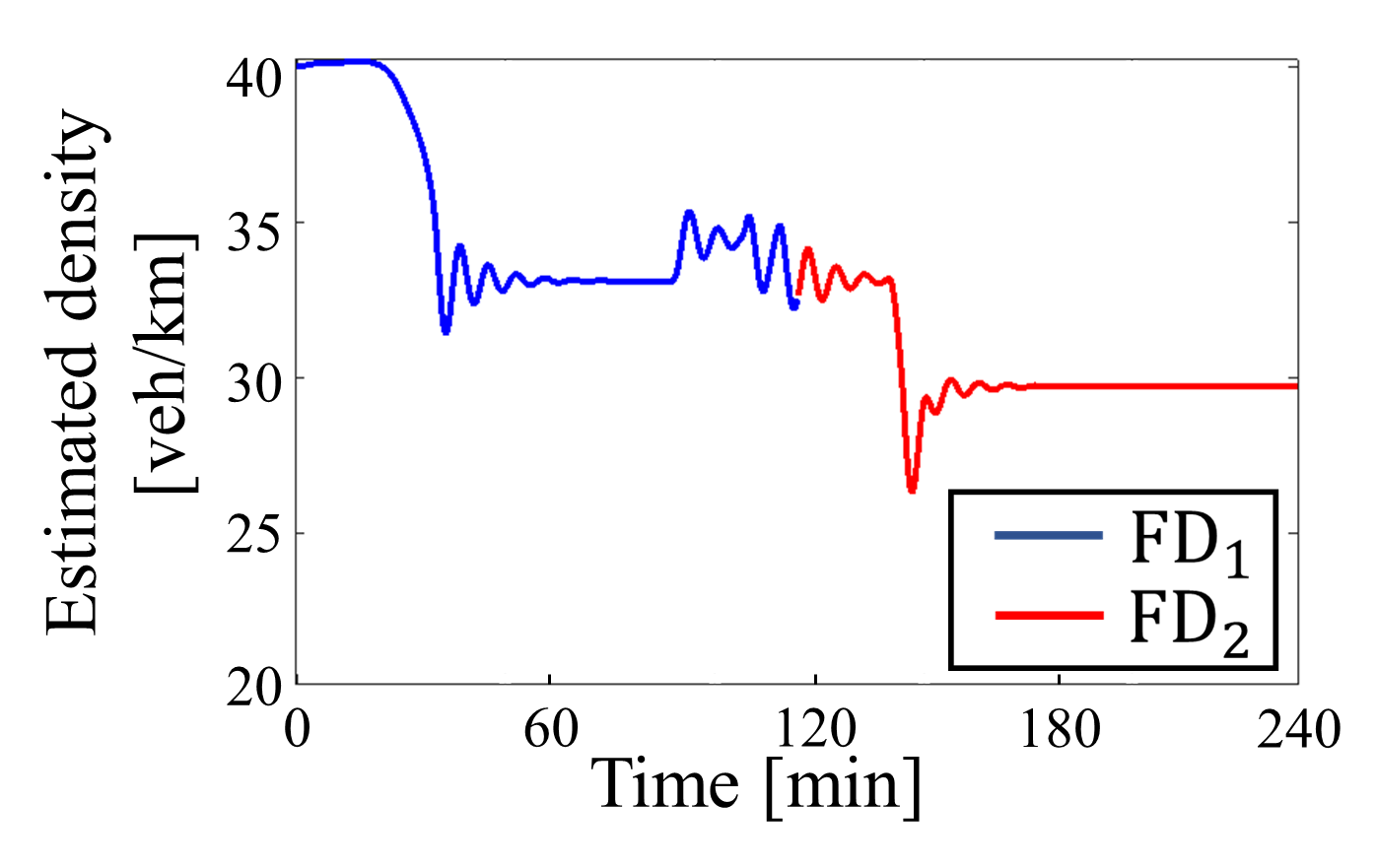}}\hspace{0cm} \\
    \subfloat[]{\includegraphics[width = 0.49\linewidth]{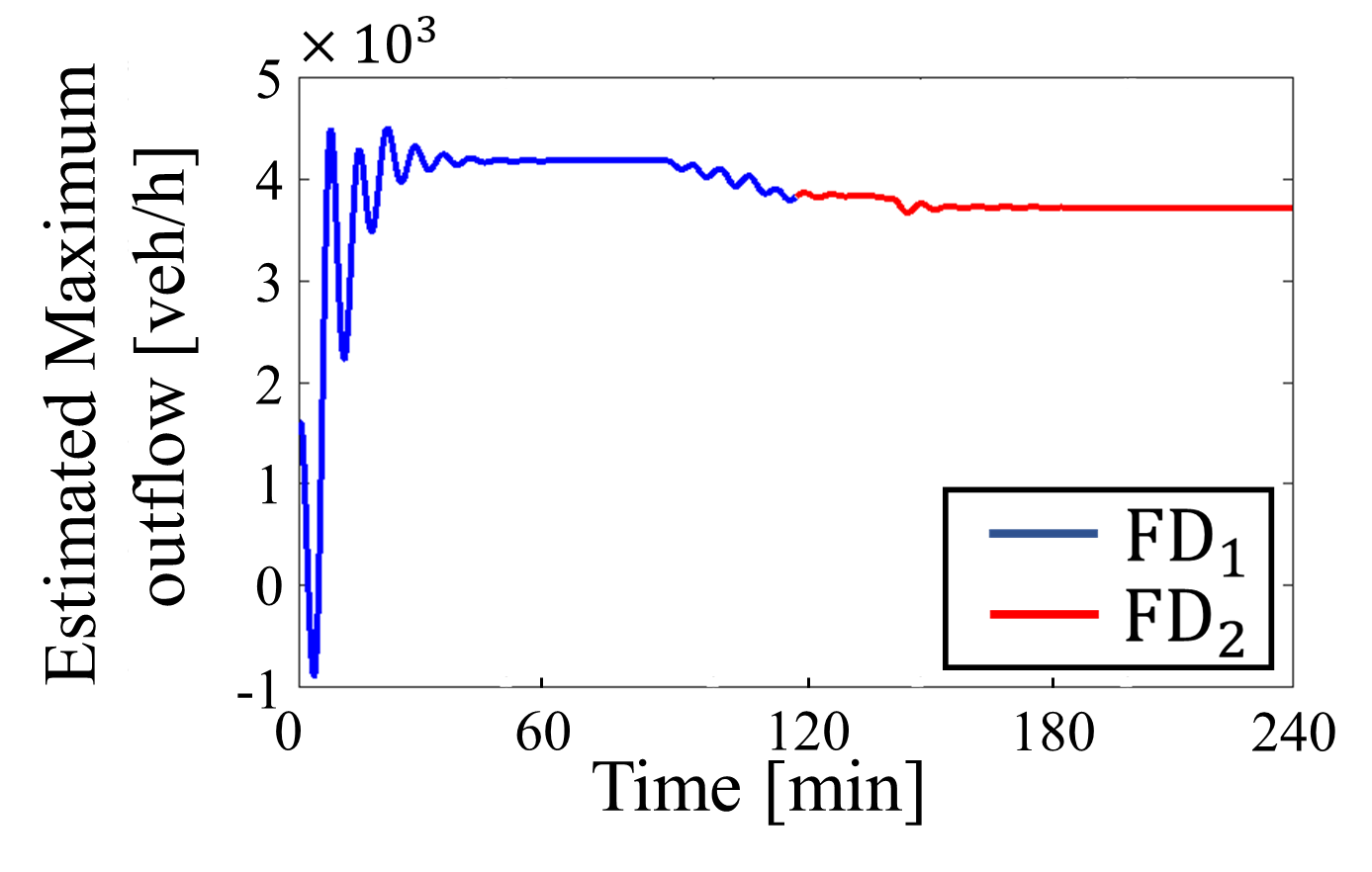}}\hspace{0cm}
    \subfloat[]{\includegraphics[width = 0.49\linewidth]{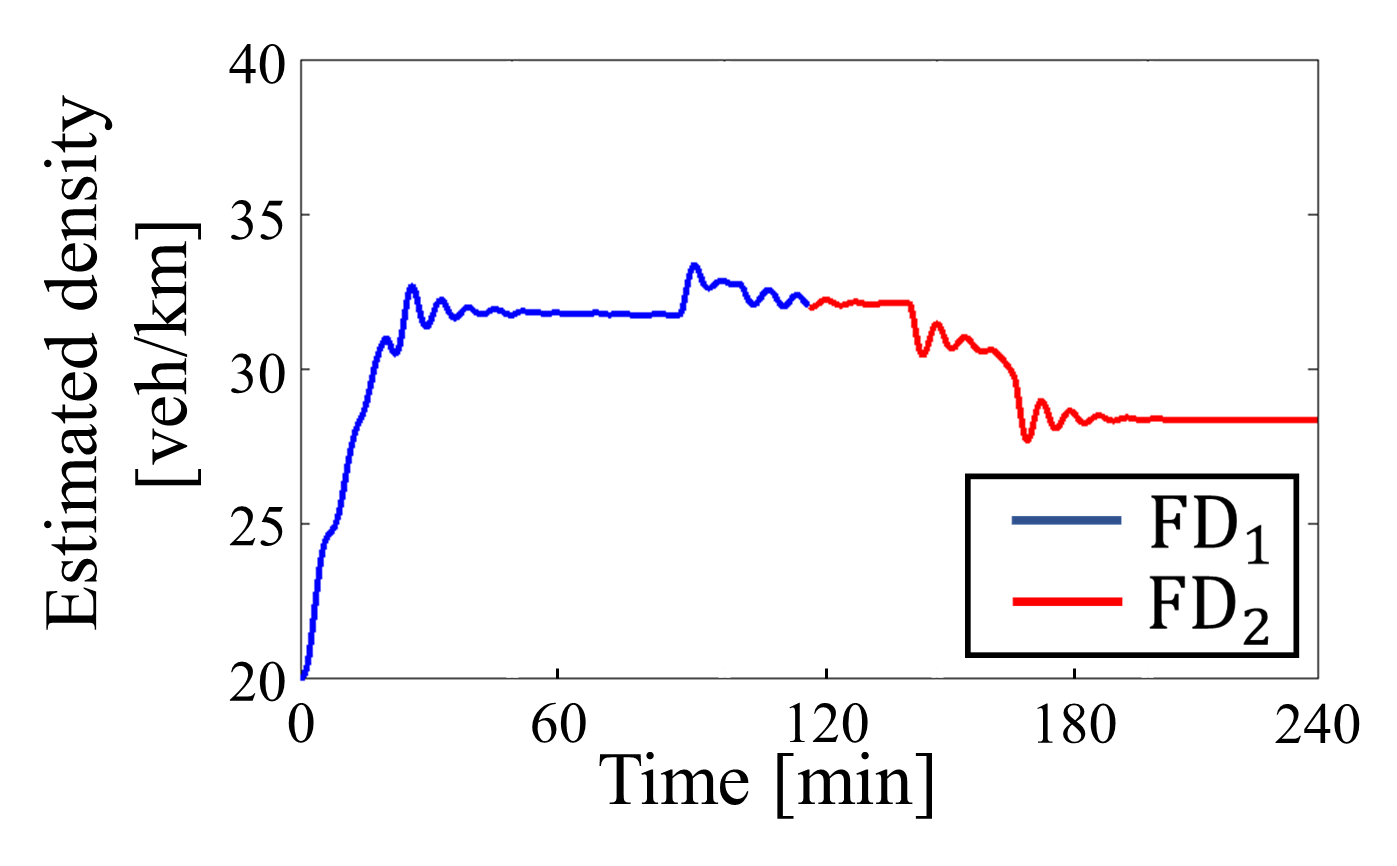}}\hspace{0cm} \\
	\caption{Estimated values, left: maximum out flow ($q^{\star}$), right: Critical density ($\rho^{\star}$) for (a,b) Scenario 5-a, where $\hat{\rho}^{\star}(0) = 40$ veh/km. (c,d) Scenario 5-b, where $\hat{\rho}^{\star}(0) = 20$ veh/km.}
	\label{fig:deviceExtreme}
\end{figure}

Looking at Figs~\ref{fig:DensityPerformExtreme}(a,b), we observe that the densities at the bottleneck area are maintained around their true critical values for both Scenarios~5-a and~5-b, although the initial conditions are considerably far from the actual values. 
This demonstrates that the proposed estimator is capable of achieving a proper convergence in a short time. 
In particular, we can see from Figs.~\ref{fig:deviceExtreme}(b,d) that, for both Scenarios~5-a and~5-b, the estimator reaches the actual set-point value ($\rho^{\star}~=~33$~veh/km) around $t = 25$ min, while Figs.~\ref{fig:deviceExtreme}(a,c) show that the maximum outflow ($\hat{q}^{\star}$) is also properly estimated. 

Also for this Scenario, the TTS values are reported in Table~\ref{tab:TTS}, where we can see that the percentage of TTS improvement is about 3.5\% and 3.3\%, respectively, compared to the no-control case (Scenario~1), thus outperforming Scenario~3.

\subsection{Sensitivity analysis of the reference model parameters}
\label{subsec:Sensitivity}
Although the reference model defined in Section~\ref{subsec:ReferenceModel} is proven to be globally stable, which guarantees the convergence of the estimated parameters, the quality and speed of the estimation process may be affected by a proper choice of the parameters $K_r$ and $C_r$. In order to investigate their effect, we perform a set of experiments considering Scenario 4, as introduced in Section~\ref{sec:Scenario4}, and comparing the resulting TTS to better understand the sensitivity of the convergence process and tracking error to the choice of such parameters. 
\begin{figure}[tb]
\centering
	 \subfloat{\includegraphics[width = 0.5\linewidth]{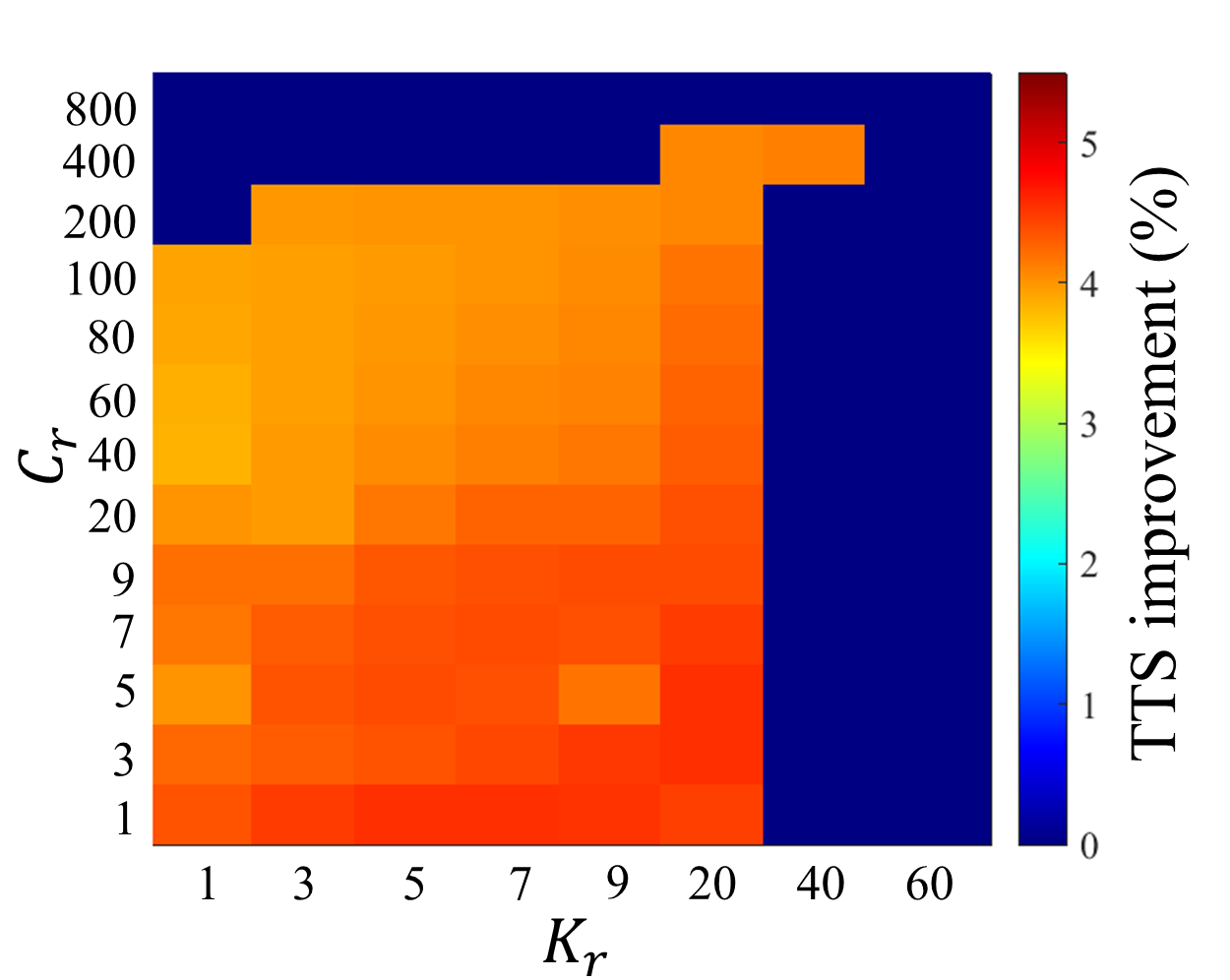}}\hspace{0 cm}
	\caption{Sensitivity analysis showing the percentage of TTS improvement compared to the no-control case for a domain of $K_r$ and $C_r$.}
	\label{fig:RMsensitivity}
\end{figure}
The results are reported in Fig.~\ref{fig:RMsensitivity}, where one may observe that the ranges of $K_r$ and $C_r$ that produce positive effects in terms of TTS improvement (i.e., the orange area) is very wide, that is, the estimator is not very sensitive to such parameters choice while we remain within these ranges.
Still, one may observe a darker orange area, where $1 \leq K_r \leq 20$, and $1~\leq~C_r~\leq 9$, which leads to the best performance in terms of TTS improvement. Thus, for our experiments, we select $K_r=10$, and $C_r=2$, which lie in this area.

\section{Conclusions}
\label{sec:conclution}

This paper proposed a novel robust adaptive estimator to estimate the  set-point values (i.e., critical density) for local traffic control strategies, designed to achieve maximum throughput at a bottleneck area, assuming the FD is unknown and time-varying.
The global asymptotical stability of the estimator is proven through a Lyapunov function, guaranteeing convergence to the true critical density and maximum outflow. In addition, the stability and convergence of the estimator's parameters are investigated via a least-square method. 
We implemented the controller and the estimator with the feedback controller for ramp metering ALINEA, utilising the traffic flow model METANET modified to account for  a time-varying FD.
Our numerical results show that employing the adaptive estimator outperforms, in terms of TTS, the ALINEA controller in the case a constant set-point is utilised. Furthermore, to assess the robustness of the estimator, we tested extreme cases for the initial estimates.  

Further developments include the incorporation in the control strategy of mainstream flow control, which may be implemented, for example, via variable speed limits, as well as accounting for the presence of multiple bottlenecks; the latter could, e.g., follow the works in \cite{wang2010local,Iordanidou2014a}. Another possible direction is to investigate the case of more complex networks, characterised by multiple destinations, where, e.g., the behaviour of CAVs is defined per destination.


\section*{Acknowledgments}
The research leading to these results has received funding from Academy of Finland projects ULTRA and AIforLEssAuto, as well as the FINEST Twins Center of Excellence (H2020 grant agreement No 856602).

\bibliographystyle{IEEEtran}
\bibliography{IEEEabrv,mybibfile}

\begin{IEEEbiography}[{\includegraphics[width=1in,height=1.25in,clip,keepaspectratio]{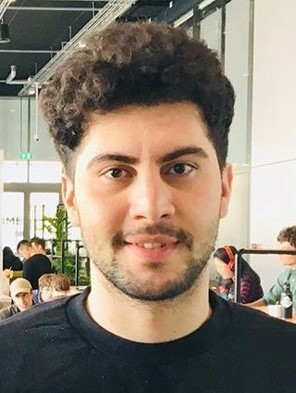}}]{Farzam Tajdari}
is a PhD student at Aalto University, Finland. He received his master's degree in Mechanical Engineering in 2016, from Amirkabir University of Technology, Tehran, Iran. Currently, He is working on autonomous and connected vehicles. His research interests include control, optimisation, and non-linear systems to solve challenges in the fields of intelligent transportation systems and geometry processing. 
\end{IEEEbiography}

\begin{IEEEbiography}[{\includegraphics[width=1in,height=1.25in,clip,keepaspectratio]{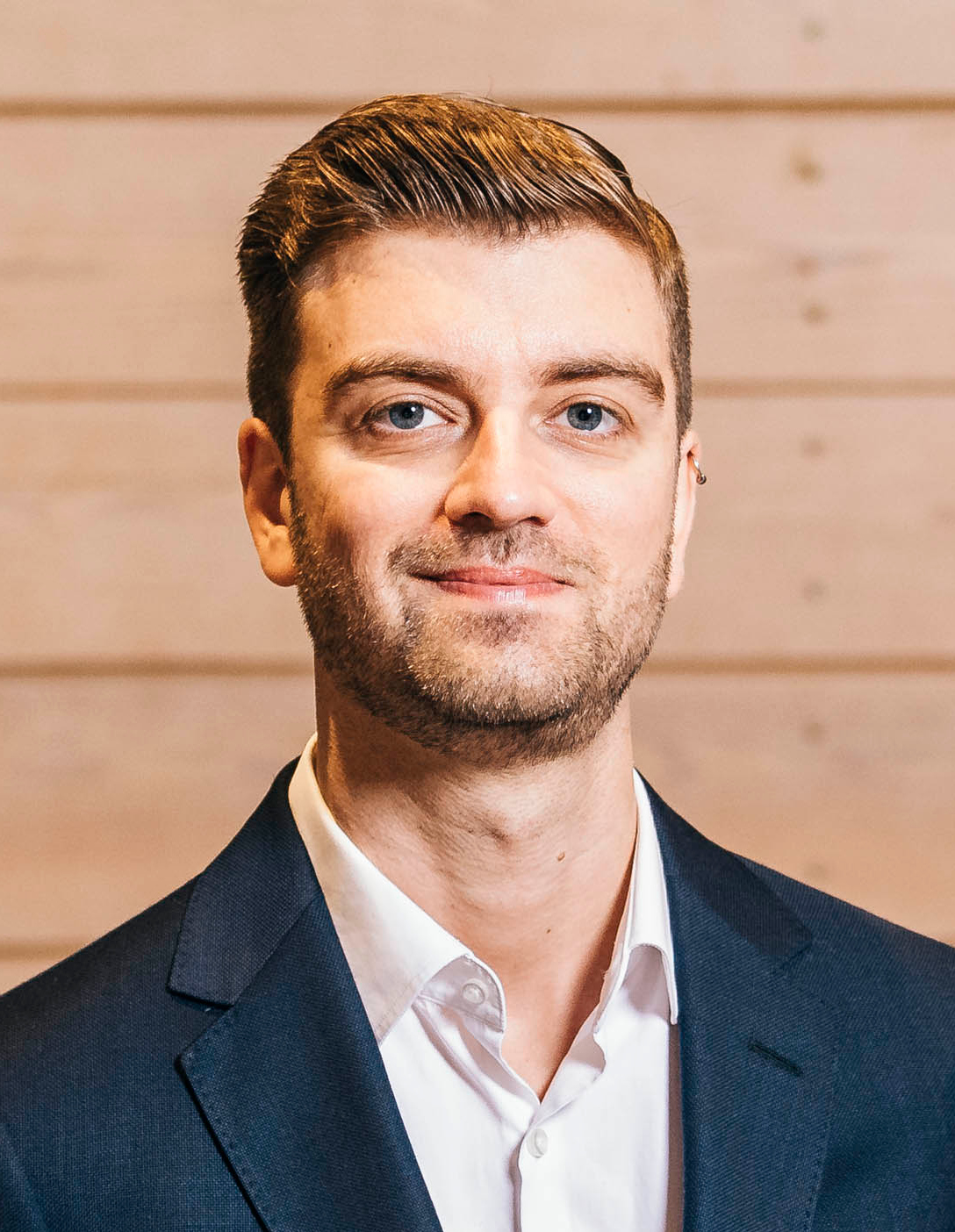}}]{Claudio Roncoli}
is Assistant Professor of Transportation Engineering at Aalto University, Finland. He completed his PhD degree in 2013 at the University of Genova, Italy.
Before joining Aalto University, he was a research assistant at the University of Genova, Italy, a visiting research assistant at Imperial College London, UK, and a Postdoctoral Researcher at the Technical University of Crete, Greece.
Claudio has been involved in several national and international research projects as a principal investigator. His research interests include real-time traffic management; modelling, optimisation, and control of traffic systems with connected and automated vehicles; as well as smart mobility and intelligent transportation systems.
\end{IEEEbiography}


\vfill

\end{document}

%% file: tabParFD.tex
\begin{tabular}{p{0.25cm} p{0.7cm} p{0.7cm} p{0.7cm} p{0.7cm} p{0.4cm} p{0.15cm} p{0.15cm} p{0.15cm} p{0.7cm}}
	\hline\hline
	& $v^{\textrm{max-min}}$ & $Q^{\textrm{cap}}$ & $\rho^{\textrm{cr}}$ & $\rho^{\textrm{jam}}$ & $\tau$ & $\nu$ & $\kappa$ & $\delta$ & $\alpha$ \\
	& [km/h] & [veh/h] & [veh/km] & [veh/km] & & & & \\
	\hline
	FD$_1$ & 107-7 & 2000 & 29 & 180 & $\frac{20}{3600}$ & 35 & 13 & 0.8 & 2.2768\\
	FD$_2$ & 107-7 & 1800 & 26 & 180 & $\frac{20}{3600}$ & 35 & 13 & 0.8 & 2.2968\\
	\hline\hline
\end{tabular}

%% file: TTS.tex
\begin{tabular}{p{1.3cm} p{1.3cm} p{1cm} p{1.2cm} p{1.5cm}}
	\hline\hline
	Case study & $\hat{\rho}^{\star}(0)$ [veh/km] & Estimator & TTS $\textrm{[veh}$$\cdot$$\textrm{h]}$ & Improvement $(\%)$\\ 
	\hline \hline
	Scenario~1& $-$ & $-$ & 1741 & $-$\\ 
	\hline 
	Scenario~2& $-$ & Off &  1638 & 5.9 \\ 
	\hline 
	\multirow{2}{*}{Scenario~3}   & (a):~33& Off & 1708 & 1.9 \\ \cline{2-5}
	                              & (b):~28& Off & 1695 & 2.5 \\ 
    \hline 
	\multirow{2}{*}{Scenario~4}   & (a):~33& On & 1687 & 3.1 \\ \cline{2-5}
	                              & (b):~28& On & 1647 & 5.4 \\ 

    \hline 
	\multirow{2}{*}{Scenario~5} & (a):~40 & On & 1680 & 3.5 \\ \cline{2-5}
	                              & (b):~20& On & 1684 & 3.3 \\ \cline{2-5}
	\hline \hline 
\end{tabular}
